\documentclass[epj]{svjour}
\usepackage{epsfig}

\newcommand{\be}{\begin{equation}}
\newcommand{\ee}{\end{equation}}
\newcommand{\bea}{\begin{eqnarray}}
\newcommand{\eea}{\end{eqnarray}}

\newcommand{\mel}[3]{\langle #1\,|\,#2\,|\,#3\rangle}
\newcommand{\scp}[2]{\mbox{$\langle #1\,|\,#2\rangle$}}
\newcommand{\scpBig}[2]{\Big\langle #1\,\Big|\,#2\Big\rangle}
\newcommand{\av}[1]{\mbox{$\langle \, #1 \, \rangle$}}

\newcommand{\ket}[1]{\mbox{$|#1\rangle$}}
\newcommand{\sep}[2]{\mbox{$|\, #1\rangle\langle #2 \, |$}}

\renewcommand\makeheadbox{}

\begin{document}

\title{Bayesian Reconstruction of Approximately Periodic Potentials
at Finite Temperature}
\author{J. C.  Lemm\thanks{e-mail: {\tt lemm@uni-muenster.de}}, 
        J. Uhlig, A. Weiguny}
\authorrunning{J. C. Lemm {\it et al.}}
\titlerunning{Bayesian Reconstruction of Approximately Periodic Potentials
at Finite Temperature}
\institute{
Institut f\"ur Theoretische Physik,\\
Universit\"at M\"unster, 48149 M\"unster, Germany}

\abstract{
The paper discusses the reconstruction of potentials 
for quantum systems at finite temperatures
from observational data.
A nonparametric approach is developed, 
based on the framework of Bayesian statistics,
to solve such inverse problems.
Besides the specific model of quantum statistics
giving the probability of observational data,
a Bayesian approach is essentially based
on {\it a priori} information available for the potential.
Different possibilities to implement
{\it a priori} information
are discussed in detail, including hyperparameters,
hyperfields, and non--Gaussian auxiliary fields.
Special emphasis is put on the reconstruction
of potentials with approximate periodicity. 
The feasibility of the approach
is demonstrated for a numerical model.
}

\date{\today}
\maketitle

\tableofcontents

\section{Introduction}

A successful application of quantum mechanics to real world systems
relies essentially on an adequate reconstruction of the underlying potential,
describing the forces governing the system.
The reconstruction of potentials or forces
from available observational data
defines an empirical learning task.
It also constitutes a typical example of an inverse problem.
Such problems are notoriously ill--defined in the sense of Tikhonov
\cite{Tikhonov-Arsenin-1977,Kirsch-1996,Vapnik-1998,Honerkamp-1998}.
In that case additional {\it a priori} information
is required to yield a unique and stable solution.
A Bayesian framework 
is especially well suited to include both, observational data and 
{\it a priori} information, in a quite flexible manner.

Inverse scattering theory
\cite{Newton-1989,Chadan-Sabatier-1989,Chadan-Colton-Paivarinta-Rundell-1997} 
and inverse spectral theory 
\cite{Gelfand-Levitan-1951,Kac-1966,Marchenko-1986,Zakhariev-Chabanov-1997}
are two classical research fields
which deal in particular with the reconstruction of potentials
from spectral data.
Both theories describe the kind of data 
which are necessary, in addition to a given spectrum, 
to determine a potential uniquely.
In inverse scattering theory these additional data
are for example phase shifts, obtained far away from the scatterer.
For the bound state problems studied in inverse spectral theory 
these additional data may consist of a second spectrum
obtained for boundary conditions different from 
those for the first spectrum.
The approach of Bayesian Inverse Quantum Mechanics (BIQM)
we will refer to in the following 
is not exclusively designed for spectral data
but is able to work with quite arbitrary observational data
\cite{Lemm-IQS-2000}.
It can thus be easily adapted to a large variety of
different reconstruction scenarios 
\cite{Lemm-BFT-1999,Lemm-TDQ-2000,Lemm-IHF-2000}.

The basics of a Bayesian framework are summarized in Section \ref{bayesian}.
Setting up a Bayesian approach for a specific application area
requires the definition of two basic probabilistic models.
First, a {\it likelihood model} is needed
giving, for each possible potential,
the probability of the observational data.
The likelihood model of quantum statistics
is discussed in Section \ref{Likelihood-model}.
Second, 
a {\it prior model} has to be chosen to implement available 
{\it a priori} information.
Prior models
which are useful for inverse quantum statistics
are presented in Section \ref{Prior-models}.
Technically the most convenient prior models
are Gaussian processes, presented in Section \ref{Gaussian-processes}.
Section \ref{Covariances-and-approximate-symmetries}
shows how covariance and mean of a Gaussian process
can be related to {\it a priori information}
about approximate symmetries of the potentials to be reconstructed.
Section \ref{Approximate-periodicity}
concentrates on approximate periodicity,
Section \ref{discontinuities} on potentials with discontinuities.
Prior models are made more flexible by using {\it hyperparameters}
(Section \ref{hyperparameter}), or more general
{\it hyperfields},
being function hyperparameters (Section \ref{hyperfields}).
Related non--Gaussian priors
are the topic of Section \ref{Non--Gaussian-priors}.
Having defined liklihood and prior models
Section \ref{stationarity-equations}
discusses the equations to be solved
for reconstructing a potential. 
Finally, 
Section \ref{numerical}
presents numerical applications.

\section{Bayesian approach}
\label{bayesian}

Empirical learning is based on observational data $D$.
In particular, we will distinguish ``dependent'' variables $x$,
representing measurement results,
and ``independent'' variables $O$,
characterizing the kind of measurement performed.
In the context of inverse quantum mechanics 
the latter denotes the {\it observables} which are measured.
Such observables may for example be 
the position, the momentum, or the energy of a quantum particle.
Variables $x$ and $O$ are assumed to be measurable
and represent therefore {\it visible} variables.
Observational data will be assumed to consist of $n$ pairs
$D$ = $\{(x_i,O_i)|1\le i\le n\}$ = $(x_T,O_T)$,
where $x_T$ and $O_T$ denote the vectors
with components $x_i$ or $O_i$, respectively.
Such data will also be called {\it training data}.
In empirical learning one tries to extract a 
``general law'' from observations.
In this paper the quantum potential $V$ 
to be reconstructed will represent this ``general law''.
(Similarly, in the Bayesian reconstruction of quantum states
the object to be reconstructed is the density operator
of an unknown state 
\cite{Helstrom:1976,Holevo:1982,Tan:1997,Buzek-Drobny-Derka-Adam-Wiedemann:1998}.)
Potentials, considered not to be directly observable,
represent in our context the
{\it hidden} or {\it latent} variables.
We will now use the Bayesian framework to relate
unobservable potentials to observational data.

The Bayesian approach is a general probabilistic framework
to deal with empirical learning problems 
\cite{Bayes-1763,Berger-1980,Loredo-1990,Bernado-Smith-1994,Gelman-Carlin-Stern-Rubin-1995,Sivia-1996,Carlin-Louis-1996,Lemm-BFT-1999}.
Predicting results of future measurements 
on the basis of given training data
is achieved by means of 
the {\it predictive probability} 
$p(x|O,D)$
(or predictive density for continuous $x$), 
which is the probability 
of finding the value $x$ when measuring observable $O$
under the condition that the training data $D$ are given.
To calculate the predictive
probability a probabilistic model 
is needed
which describes the measurement process.
Such a model is specified by
giving  the probability  $p(x|O,V)$
of finding $x$ when measuring observable $O$
for each possible potential $V$.
As $p(x|O,V)$, considered as function of $V$ for fixed $x$ and $O$,
is known as likelihood of $V$,
we will call this the {\it likelihood model}.
For inverse quantum problems
the likelihood model is given by the axioms of quantum mechanics
and will be discussed in Section \ref{Likelihood-model}.

According to the rules of probability theory
the predictive probability can now be written as an integral
over the space of all possible potentials $V$,
\be
p(x|O,D)
= \int \!dV\, p(x|O,V)\, p(V|D)
.
\label{predictive}
\ee 
We note that in Eq.(\ref{predictive}) we have assumed
that the probability of $x$ is completely determined 
by giving potential and observable
and does not depend on the training data, $p(x|O,V,D)$ = $p(x|O,V)$,
and
that the probability of the potential given the training data
does not depend on the observables selected in the future,
$p(V|O,D)$ = $p(V|D)$.
If the set of possible potentials is a space of functions, 
the integral in (\ref{predictive}) is a functional integral.

As the likelihood model is assumed to be given,
learning consists in the determination of $p(V|D)$,
known as the {\it posterior} for $V$.
To this end, we relate the 
posterior for $V$ to the
likelihood of $V$ under the training data
by applying Bayes' theorem,
\be
p(V|D)
=
\frac{p(x_T|O_T,V)\,p(V)}{p(x_T|O_T)}
,
\label{bayestheorem}
\ee
assuming $p(V|O_T)$ = $p(V)$,
analogous to Eq.~(\ref{predictive}).
In the numerator of Eq.~(\ref{bayestheorem})
appears, besides the likelihood, 
the so called prior $p(V)$.
This prior gives the probability of $V$ 
{\it before} training data have been collected.
Hence it has to comprise all 
{\it a priori information}
available for the potential.
The need for a prior model,
complementing the likelihood model,
is characteristic for a Bayesian approach.
The denominator in Eq.~(\ref{bayestheorem})
plays the role of a normalization factor
and can be obtained from likelihood and prior
by integration over $V$ as
$p(x_T|O_T)$ 
= $\int \!dV\,p(x_T|O_T,V)\,p(V)$.

From a Bayesian perspective learning appears as
updating the probability for $V$ caused by the arrival of new data $D$.
If more data become available
this process can be iterated, 
the old posterior becoming the new prior
which is then updated yielding a new posterior.

In practice, a major difficulty is the calculation of 
the integral over all possible $V$
to get the predictive probability (\ref{predictive}). 
Even if one resorts to a discrete approximation for $x$
the integral (\ref{predictive}) is
typically still very high dimensional.
The key point is thus to find a feasible
approximation for  that integral.
Two approaches are common in Bayesian statistics.
The first one is an evaluation of the integral
by Monte Carlo methods 
\cite{Gelman-Carlin-Stern-Rubin-1995,Metropolis-Rosenbluth-Rosenbluth-Teller-Teller-1953,Binder-Heermann-1988,Neal-1997}.
The second one, which we will pursue in the following, 
is the so called {\it maximum a posteriori approximation} (MAP),
being a variant of the saddle point method
\cite{Berger-1980,Gelman-Carlin-Stern-Rubin-1995,De-Bruijn-1981,Bleistein-Handelsman-1986,Girosi-Jones-Poggio-1995,Lemm-1996,Lemm-1998}.
In MAP one assumes the posterior to be sufficiently peaked
around the potential $V^*$ which maximizes the posterior,
so that approximately
\be
p(x|O,D) \approx p(x|O,V^*)
,
\ee
with 
\be
V^* 
= {\rm argmax}_{V\in{\cal V}} p(V|D)
= {\rm argmax}_{V\in{\cal V}} p(x_T|O_T,V) p(V)
.
\label{map-eq}
\ee
Maximizing the posterior with respect to $V\in{\cal V}$
means,  according to Eq.~(\ref{bayestheorem})
with the denominator independent of $V$,
maximizing the product of likelihood and prior.

The Bayesian framework discussed
so far can analogously be applied to a variety of different contexts,
including regression, density estimation and classification
problems \cite{Lemm-BFT-1999}.
The case of a Gaussian likelihood with fixed variance, for example,
is known as regression problem,
while problems with general likelihoods
are known as density estimation.

\section{Likelihood model of quantum statistics}
\label{Likelihood-model}

The first step in applying the Bayesian framework
to inverse problems of quantum mechanics or quantum statistics
is the definition of the likelihood model \cite{Lemm-IQS-2000}.
This is easily obtained from the axioms of quantum mechanics.
Consider a system prepared in a state described by
a density operator $\rho$. 
As our aim will be to reconstruct potentials $V$
from observational data,
we have to choose a $\rho$ which depends on the potential.
The probability to find value $x$,
when measuring an observable represented by the Hermitian operator $O$,
is given by
\be
p(x|O,V) 
= {\rm Tr} 
\Big(P_O(x) \, \rho(V) \Big)
,
\label{qm-likelihood}
\ee
where 
$P_O(x)$ = $\sum_\zeta \sep{x,\zeta}{x,\zeta}$
denotes the projector on the space of 
(orthonormalized) eigenfunctions $\ket{x,\zeta}$ 
of $O$ with eigenvalue $x$ and
the variable $\zeta$ distinguishes 
eigenfunctions with degenerate eigenvalues.

In particular, for a canonical ensemble
at temperature $1/\beta$ 
(setting  Boltzmann's constant equal to 1)
the density operator reads
\be
\rho = 
\frac{e^{-\beta H}}{{\rm Tr\,} e^{-\beta H}}
.
\label{canonical}
\ee
To be specific, we will study in the following Hamiltonians of the form
$H$ = $T + V$,
with kinetic energy 
$T$ = $-(1/2m)\Delta$,
(with Laplacian $\Delta$, mass $m$, 
and setting $\hbar$ = $1$)
and a 
local potential
\be
V(x,x^\prime) = 
v(x) \delta (x-x^\prime )
,
\ee
defined by the function $v(x)$.
Note that the formalism presented in the following 
works with nonlocal potentials as well,
numerical calculations, however, would in that case be more
demanding.
For the likelihood models corresponding to
time--dependent quantum systems
and to many--body systems 
in Hartree--Fock approximation
we refer to \cite{Lemm-TDQ-2000,Lemm-IHF-2000}.

In the following we will study observational data
consisting of $n$ position measurements $x_i$.
This corresponds to choosing the position operator 
for the observables $O_i$ = $\hat x$
with 
$\hat x \ket{x_i}$ = $x_i\ket{x_i}$.
Hence, for a canonical ensemble, 
the likelihood (\ref{qm-likelihood})
becomes
for a single position measurement
\be
p(x_i|\hat x,v) 
=\sum_\alpha p_\alpha |{\phi}_\alpha(x_i)|^2 
=\av{|{\phi} (x_i)|^2}
\label{pos-likelihood}
\ee
with (non--degenerate) eigenfunctions ${\phi}_\alpha$ of $H$
and energies $E_\alpha$,
i.e.,
$H\ket{{\phi}_\alpha}$ = $E_\alpha \ket{{\phi}_\alpha}$.
Angular brackets $\av{\cdots}$
denote a thermal expectation 
under the probabilities
$p_\alpha$ = 
$\exp(-\beta E_\alpha)/Z$ with
$Z$ = $\sum_\alpha \exp (-\beta E_\alpha)$
according to Eq.~(\ref{canonical}).
For independent data $D_i$ = $(x_i,O_i)$,
\be
p(x_T|O_T,v) 
= \prod_{i=1}^n p(x_i|\hat x,v) 
= \prod_{i=1}^n \av{|{\phi} (x_i)|^2}
.
\ee
A quantum mechanical measurement changes 
the state of the system, i.e., it changes $\rho$.
Hence, to obtain independent data under constant $\rho$
requires the density operator 
to be restored before each measurement.
For a canonical ensemble this means 
to wait between two consecutive observations
until the system is thermalized again.

Choosing a parametric family of potentials
$v(x;\xi)$
one could now
maximize the likelihood
with respect to the parameters $\xi$,
and choose as reconstructed potential
\be
v^*(x) = v(x;\xi^*)
\quad \mbox{with} \quad
\xi^* 
= \mbox{\rm argmax}_{\xi} \, p(x_T|O_T,v(\xi) )
.
\label{ml-eq}
\ee
This is known as maximum likelihood approximation 
and works well 
if the number of data is large compared to the 
flexibility of the selected parametric family of potentials.
This method does however not yield a unique optimal potential
if the flexibility is too large for the available number of observations.
(A possible measure of the ``flexibility'' of a parametric family
is given by the Vapnik-Chervonenkis dimension \cite{Vapnik-1998}
or variants thereof.)
In such cases, the inclusion of additional restrictions on $v$ 
in form of {\it a priori} information is essential.
This holds especially for nonparametric approaches,
where each number $v(x)$ is treated
as individual degree of freedom.
Including {\it a priori} information
generalizes the 
maximum likelihood approximation 
of Eq.~(\ref{ml-eq})
to the MAP of Eq.~(\ref{map-eq}).

\section{Prior models}
\label{Prior-models}

\subsection{Gaussian processes}
\label{Gaussian-processes}

A finite number of observational data cannot
completely determine a function $v(x)$.
Hence,  besides observational data,
additional {\it a priori} information
is necessary to reconstruct a potential in BIQM.
In nonparametric approaches
it is advantageous to formulate
{\it a priori} information
directly in terms of the function $v(x)$ itself.
A convenient choice for a prior is a Gaussian process,
\be
p(v) 
=
\left(\det \frac{{\bf K}_0}{2\pi}\right)^\frac{1}{2}
e^{-\frac{1}{2} \mel{v-v_0}{{\bf K}_0}{v-v_0}}
,
\label{gaussprior}
\ee
where 
\be
\mel{v-v_0}{{\bf K}_0}{v-v_0}
= 
\ee
\[
\int\! dx \,dx^\prime\, [v(x)-v_0(x)]{\bf K}_0(x,x^\prime)
[v(x^\prime)-v_0(x^\prime)].
\]
The function $v_0$ is the mean or regression function,
representing a reference potential or template for $v$.
The inverse covariance ${\bf K}_0$
is a real symmetric, positive (semi)definite operator 
which acts on potentials rather than on wave functions
and defines
a distance measure on the space of potentials.
For technical convenience one may introduce explicitly
a factor $\lambda$ multiplying  ${\bf K}_0$
to balance the influence of the prior 
against the likelihood term.
A Gaussian prior as in Eq.~(\ref{gaussprior})
is already a quite flexible tool 
for implementing {\it a priori} knowledge.
A bias towards smooth
functions $v(x)$, for instance, 
can be implemented by choosing the negative Laplacian as inverse
covariance
${\bf K}_0$ = $-\Delta$.
Including higher derivatives in ${\bf K}_0$ 
would result in even smoother potentials,
in the sense that higher derivatives of $v(x)$ become continuous.
For example,
a common smoothness prior used for regression problems is 
the Radial Basis Function prior
${\bf K}_0$ = $\exp{(-{\sigma_{\rm RBF}^2}{\Delta}/2)}$
\cite{Girosi-Jones-Poggio-1995}.

\subsection{Covariances and approximate symmetries}
\label{Covariances-and-approximate-symmetries}

Prior information on potentials $v$ can often be related to 
approximate invariance under specific transformations
\cite{Lemm-BFT-1999}.
Typical examples of such transformations are symmetry operations
like translations or rotations.
To be specific, assume that
a (not necessarily local) potential $V$ 
commutes approximately, but not exactly,
with some unitary operator $S$, 
\be
V \approx S^\dagger V S = {\bf S} V
,
\ee
which defines an operator ${\bf S}$
acting on $V$.
In particular, we may choose a prior 
$p(V)\propto\exp \{-E_{S}(V)\}$ with
a {\it prior energy}
\be
E_S
= \frac{1}{2}\scp{V-{\bf S}V}{V-{\bf S}V}
= \frac{1}{2}\mel{V}{{\bf K}_{0}}{V}
.
\ee
This shows that the expectation 
of an approximate symmetry of $V$ under $S$
can be implemented by choosing a Gaussian prior with
inverse covariance operator
\be
{\bf K}_0 = 
({\bf I}-{\bf S})^\dagger ({\bf I}-{\bf S})
,
\ee
where ${\bf I}$ denotes the identity operator. 
Symmetry operations $S(\theta)$,
with corresponding ${\bf S}(\theta)$,
may depend on a parameter (vector) $\theta$.
Approximate invariance under $S(\theta_i)$ 
for several $\theta_i$
can be implemented by using the sum 
(or integral, for continuous variables)
\bea
E_S 
&=& \frac{1}{2}
\sum_i \scp{V-{\bf S}(\theta_i)V}{V-{\bf S}(\theta_i)V}
\nonumber\\
&=& \frac{1}{2}\sum_i \mel{V}{{\bf K}_{0}(\theta_i)}{V}
.
\eea
Alternatively, one may 
require approximate symmetry for only one value of $\theta$,
not fixed {\it a priori}.
For example, one may expect an approximately periodic potential
with unknown periodicity length $\theta$
which also has to be determined from the data.
Such $\theta$ are known as {\it hyperparameters}
and will be discussed in Section \ref{hyperparameter}.

Lie groups
are continuously parameterized transformations
\be
{\bf S}(\theta)
=e^{\sum_i\theta_i {\bf s}_i}
,
\ee
where $\theta_i$ are the real parameters 
and the ${\bf s}_i$ = $-{\bf s}_i^T$ 
(the superscript ${}^T$ denoting the transpose)
are antisymmetric operators
representing the generators 
of the infinitesimal transformations 
of the Lie--group.
We can define a prior energy as
an error measure with respect to an
infinitesimal transformation,
\bea
E_S &=&
\frac{1}{2}
\sum_i  
\scpBig{\frac{{V} - (1 + \theta_i {\bf s}_i) {V}}{\theta_i}}
    {{\frac{{V} - (1 + \theta_i {\bf s}_i){V}}{\theta_i}}} 
\nonumber\\
&=&
\frac{1}{2}
\mel{{V}}{\sum_i {\bf s}_i^T {\bf s}_i}{{V}}
\label{Lie-error}
.
\eea
For instance, a Laplacian smoothness prior for a local potential $v(x)$
can be related to an approximate symmetry 
under infinitesimal translations.
For the group of 
$d$--dimensional translations which is generated
by the gradient operator $\nabla$
this can be verified by recalling the multidimensional Taylor formula 
for expanding ${v}$ around $x$ 
\be
{\bf S}(\theta) {v}(x) 
= e^{ \sum_i \theta_i \nabla_i } {v}(x)
= \sum_{k=0}^\infty 
\frac{\left(\sum_i \theta_i \nabla_i\right)^{k}}{k!} {v}(x)
= {v}(x+\theta).
\ee
Up to first order 
${\bf S} \approx  1+\sum_i\theta_i \nabla_i$.
Hence, for infinitesimal translations,
the error measure of Eq.\ (\ref{Lie-error}) becomes
\bea
E_S 
&=&
\frac{1}{2}\sum_i  
\scpBig{\frac{{v} -(1 + \theta_i {\nabla}_i) {v}}{\theta_i}}
       {{\frac{{v} - (1 + \theta_i {\nabla}_i){v}}{\theta_i}}}
\nonumber\\
&=&
-\frac{1}{2}\mel{{v}}{\Delta}{{v}}
,
\eea
assuming vanishing boundary terms.
This is the classical Laplacian smoothness term.

\subsection{Approximate periodicity}
\label{Approximate-periodicity}

In this paper we will in particular be interested
in potentials which are approximately periodic.
To measure the deviation from exact periodicity
for a local potential $v(x)$ 
let us define the difference operators
\bea
\left(\nabla^{R}_\theta v\right)(x)
&=&
{v}(x+\theta)-{v}(x).
\\
\left(\nabla^{L}_\theta v\right)(x)
&=&
{v}(x)-{v}(x-\theta),
\eea
For periodic boundary conditions
$(\nabla^{L}_\theta)^T$
=
$-\nabla^{R}_\theta$,
where $(\nabla^{L}_\theta)^T$ 
denotes the transpose of $\nabla^{L}_\theta$.
Hence, the operator
\be
-\Delta_\theta 
= -\nabla^L_\theta\nabla^R_\theta
= (\nabla^R_\theta)^T\nabla^R_\theta
\ee
defined in analogy to the negative Laplacian,
is positive (semi)\-definite,
and a possible prior energy 
is an error term 
which measures the deviation from exact periodicity
for given period $\theta$,
\bea
E_S 
&=&\frac{1}{2}\int \!dx\; |{v}(x)-{v}(x+\theta)|^2
\nonumber\\
&=&
 \frac{1}{2}
\scp{\nabla^R_\theta{v}}{\nabla^R_\theta {v}}
\nonumber\\
&=&
-\frac{1}{2}
\mel{{v}}{\Delta_\theta}{{v}}
.
\label{periodic-error}
\eea
Discretizing $v$ 
the operator $\nabla^R_\theta$  
for periodic boundary conditions
becomes, 
for example on a mesh with six points and $\theta$ = $2$, 
the matrix
\be
\nabla^R_\theta  = 
\left(
\begin{tabular}{    c     c     c     c     c    c }
                   $-1$&  0  & $1$ &  0  &  0  & 0   \\
                    0  & $-1$&  0  & $1$ &  0  & 0   \\
                    0  &  0  & $-1$&  0  & $1$ & 0   \\
                    0  &  0  &  0  & $-1$&  0  & $1$ \\
                   $1$ &  0  &  0  &  0  & $-1$  & 0 \\
                    0  & $1$ &  0  &  0  &  0  & $-1$\\
\end{tabular}
\right)
,
\ee
so that
\be
-\Delta_\theta  = 
\left(
\begin{tabular}{    c     c     c     c     c    c }
                    2  &  0  & $-1$&  0  & $-1$& 0     \\
                    0  &  2  &  0  & $-1$&  0  & $-1$  \\
                   $-1$&  0  &  2  &  0  & $-1$& 0     \\
                    0  & $-1$&  0  &  2  &  0  & $-1$  \\
                   $-1$&  0  & $-1$&  0  &  2  & 0   \\
                    0  & $-1$&  0  & $-1$&  0  & 2  \\
\end{tabular}
\right)
.
\ee

As every periodic function with ${v}(x)={v}(x+\theta)$
is in the null space of $\Delta_\theta$
typically another error term has to be added 
to get a unique maximum of the posterior.
For example, combining 
a prior energy (\ref{periodic-error})
with a Laplacian smoothness term yields
a Gaussian prior of the form (\ref {gaussprior})
with inverse covariance
${\bf K}_0$ = $-\lambda (\Delta+\gamma \Delta_\theta)$
and prior energy
\be
E_S =
-\frac{\lambda}{2}
\mel{{v}}{\Delta+\gamma \Delta_\theta}{{v}}
\label{periodic-cov}
,
\ee
with weighting factors $\lambda$, $\gamma$.  
In case the period $\theta$ is not known, it can be treated
as hyperparameter as will be discussed in Section \ref{hyperparameter}.
Clearly, a nonzero reference potential $v_0$ can be included
in Eq.~(\ref{periodic-cov}).
In Eq.~(\ref{periodic-error}),
one may also sum over several periods
\be
E_S 
= \frac{1}{2} \sum_{k=1}^{k_{\rm max}} 
w(k) \int \!dx\; |{v}(x)-{v}(x+k \theta)|^2
,
\label{periodic-error2}
\ee
where $w(k)$ is a weighting function, decreasing for larger $k$.
Prior energies as in (\ref{periodic-error2})
enforce approximate periodicity
over longer distances than
a prior energy of the form (\ref{periodic-error}).
The latter, on the other hand,
is more robust than (\ref{periodic-error2})
with respect to local deviations from periodicity,
like a locally varying frequency.

Instead of choosing an
inverse covariance ${\bf K}_0$ 
with symmetric functions in its null space,
approximate symmetries can be implemented by using 
explicitly a symmetric reference function 
$v_0$ = ${\bf S} v_0$ for the Gaussian prior (\ref{gaussprior}).
For approximate periodicity, 
this would mean to choose
a periodic reference potential
$v_0(x)$ = $v_0(x+\theta)$ in the prior energy
$E_S = \frac{1}{2} \mel{{v} -v_0}{{\bf K}_0}{{v}-v_0}$
where ${\bf K}_0$ could be for example
the identity or a differential operator.
Thus a periodic reference potential 
favors a specific form for the reconstructed potential,
including a specific frequency and phase.
This is different for the covariance implementation
(\ref{periodic-error})
of approximate periodicity 
where only the frequency is relevant
and reference potentials can still be chosen arbitrarily.
They may, for example be nonperiodic functions or functions 
with even higher symmetry
like in Eq.~(\ref{periodic-cov})
where $v_0\equiv 0$ is invariant under all translations.
Flexible reference potentials will be studied in Section \ref{hyperparameter}.

\subsection{Potentials with discontinuities}
\label{discontinuities}

Smooth potentials $v(x)$ with discontinuities can either be approximated
by using discontinuous templates $v_0(x;\theta)$ 
or by eliminating matrix elements of the inverse covariance
which connect the two sides of the discontinuity.
For example, consider the discrete version 
of a negative Laplacian
with unit lattice spacing and periodic boundary conditions,
\be
{\bf K}_0 = -\Delta =
\left(
\begin{tabular}{    c     c     c     c     c    c }
                    2  & $-1$&  0  &  0  &  0  &$-1$\\
                   $-1$&  2  & $-1$&  0  &  0  & 0  \\
                    0  & $-1$&  2  & $-1$&  0  & 0  \\
                    0  &  0  & $-1$&  2  & $-1$&  0  \\
                    0  &  0  &  0  & $-1$&  2  & $-1$ \\
                   $-1$&  0  &  0  &  0  & $-1$&  2  \\
\end{tabular}
\right).
\label{discrete1}
\ee
Decomposing the matrix (\ref{discrete1})
into square roots we write ${\bf K}_0$ = ${\bf W}^T {\bf W}$ 
(see also Section \ref{hyperfields})
where a possible square root is
\be
{\bf W} = \nabla_1^R =
\left(
\begin{tabular}{    c     c     c     c     c    c }
                  $-1$ & $1$ &  0  &  0  &  0  & 0   \\
                    0  & $-1$& $1$ &  0  &  0  & 0   \\
                    0  &  0  & $-1$& $1$ &  0  &  0  \\
                    0  &  0  &  0  & $-1$& $1$ &  0  \\
                    0  &  0  &  0  &  0  & $-1$& $1$ \\
                   $1$ &  0  &  0  &  0  &  0  & $-1$\\
\end{tabular}
\right)
\label{discrete2}
.
\ee
Similarly, the derivative operator $\partial/\partial x$
represents a square root of the negative Laplacian
for periodic boundary conditions.
Two regions can now be disconnected by deleting all lines of ${\bf W}$
which have matrix elements in both regions. 
For instance, the first three points in the six--dimensional space
of Eq.~(\ref{discrete2})
can be disconnected from the last three points
by setting 
${\bf W}(3,\cdot )$ and ${\bf W}(6,\cdot )$ to zero,
\be
\tilde {\bf W} = 
\left(
\begin{tabular}{    c     c     c  |  c     c    c }
                   $-1$& $1$ &  0  &  0  &  0  & 0  \\
                    0  & $-1$& $1$ &  0  &  0  & 0  \\
                    0  &  0  &  0  &  0  &  0  & 0  \\
\hline
                    0  &  0  &  0  & $-1$& $1$ &  0  \\
                    0  &  0  &  0  &  0  & $-1$& $1$ \\
                    0  &  0  &  0  &  0  &  0  &  0  \\
\end{tabular}
\right)
\label{discrete3}
.
\ee
Squaring of $\tilde {\bf W}$ yields a positive semidefinite operator 
\be
\tilde {\bf K}_0 = {\tilde {\bf W}}^T \tilde {\bf W} =
\left(
\begin{tabular}{    c     c     c  |  c     c    c }
                    1  & $-1$&  0  &  0  &  0  & 0  \\
                  $-1$ &  2  & $-1$&  0  &  0  & 0  \\
                    0  & $-1$&  1  &  0  &  0  & 0  \\
\hline
                    0  &  0  &  0  &  1  & $-1$&  0  \\
                    0  &  0  &  0  & $-1$&  2  & $-1$ \\
                    0  &  0  &  0  &  0  & $-1$&  1  \\
\end{tabular}
\right)
\label{discrete4}
\ee
resulting in a smoothness prior which is ineffective
between points from different regions.
In contrast to using discontinuous templates,
the height of the jump at the discontinuity 
has not to be given in advance
when working with
disconnected Laplacians (or other disconnected inverse covariances).
On the other hand
training data are then required for all separated regions
to determine the free constants 
which correspond to the zero modes of the local Laplacians.
The reconstruction of discontinuous functions 
with non--Gaussian priors will be discussed in
Section \ref{Non--Gaussian-priors}.

\subsection{Hyperparameters}
\label{hyperparameter}

Parameters of the prior are known as {\it hyperparameters}
\cite{Lemm-BFT-1999,Carlin-Louis-1996,Bishop-1995b}.
Like potentials $v$, hyperparameters $\theta$  
are not directly observable and
represent hidden variables.
In the presence of hyperparameters
a prior for $v$ can be decomposed as follows
\be
p(v)
= 
\int \!d\theta\, p(v|\theta) \, p(\theta)
,
\label{theta-integral}
\ee
where $p(\theta)$ is known as {\it hyperprior}.
The likelihood does not depend on $\theta$,
the predictive probability (\ref{predictive}), 
however, 
contains then an integral over $\theta$, 
\be
p(x|O,D) =
\label{hyper-predictive}
\ee
\[\frac{1}{p(x_T|O_T)}\int \!dv\,d\theta\,  p(x|O,v)\, 
p(x_T|O_T,v)\, p(v|\theta) \, p(\theta)
.
\]
Like the integral over $v$,
the integral over $\theta$
can be calculated either by Monte Carlo methods
or in MAP.
We remark that, when a $\theta$--dependent prior
is written in terms of a corresponding prior energy
$p(v|\theta)\propto e^{-E(v|\theta)}$,
the normalization $\int\!dv\, e^{-E(v|\theta)}$
is independent of $v$ but does in general depend on $\theta$.

Hyperparameters $\theta$ can be 
single numbers or vectors.
They can describe continuous transformations,
like translation, rotation or scaling of template functions
and scaling of inverse covariance operators.
For real $\theta$ and differentiable posterior,
stationarity conditions can be found by differentiating
the posterior with respect to $\theta$.

Instead of continuous transformations
of templates or inverse covariances
one can consider
a finite collection of 
alternative reference potentials $v_i$
or alternative inverse covariances ${\bf K}_i$.
For example, a potential to be reconstructed 
may be expected to be similar to one reference potential
out of a small number of possible alternatives $v_i$.
The ``class'' variables $i$ are then
nothing else but hyperparameters 
$\theta$ with integer values.

Binary parameters allow to select from two reference functions 
or two inverse covariances
that one which fits the data best.
Indeed, writing
\bea
v_0(\theta) &=& (1-\theta) v_1 + \theta v_2,
\label{integer-hyper-t}\\
{\bf K}_0(\theta) &=& (1-\theta) {\bf K}_1 + \theta {\bf K}_2
\label{integer-hyper-K}
,
\eea
a binary $\theta\in \{0,1\}$ implements
hard switching between alternative templates or inverse covariances,
corresponding to a conditional prior
\be
p(v|\theta) \propto e^{-(1-\theta)E_1(v)-\theta E_2(v)}
\label{mix-prior-bin}
\ee
with
\bea
E_1(v)  &=& \frac{1}{2}\mel{v-v_1}{{\bf K}_1}{v-v_1}
,
\\
E_2(v) &=& \frac{1}{2}\mel{v-v_2}{{\bf K}_2}{v-v_2}
.
\eea
Similarly, a real $\theta\in [0,1]$ 
in (\ref{integer-hyper-t}) or 
(\ref{integer-hyper-K})
yields soft mixing.
In that case, however,
the mixing of templates in (\ref{integer-hyper-t})
is not equivalent to
a mixing of prior energies 
as in (\ref{mix-prior-bin})
because for real $\theta$ 
Eqs.~(\ref{integer-hyper-t})
and (\ref{integer-hyper-K})
lead to mixed terms, 
like 
$(1-\theta)\theta \mel{v-v_1}{{\bf K}_0}{v-v_2}/2$
for ${\bf K}_1$ = ${\bf K}_2$.
When $\theta$ takes integer values the integral
$\int\! d\theta$ 
becomes a sum $\sum_\theta$ 
so that prior, posterior, and predictive probability
have the form of a {\it finite mixture} 
with components $\theta$ \cite{lemm-mixture-1999}.

For a moderate number of components
one may be able to include 
all of the mixture components in the calculations.
If the number of mixture components is too large
one must select some of the components,
for example by creating a random sample
using Monte Carlo methods, 
or by solving for the $\theta^*$
with maximal posterior.
In contrast to typical optimization problems for real variables,
the corresponding integer optimization problems
are usually not very smooth with respect to $\theta$
(with smoothness defined in terms of differences instead of derivatives),
and are therefore often much harder to solve.

There exists
a variety of deterministic and stochastic integer optimization algorithms,
which may be combined with ensemble methods like genetic algorithms
\cite{Holland-1975,Goldberg-1989,Michalewicz-1992,Schwefel-1995,Mitchell-1996},
and with homotopy methods like simulated annealing
\cite{Kirkpatrick-Gelatt-Vecchi-1983,Mezard-Parisi-Virasoro-1987,Aarts-Korts-1989,Gelfand-Mitter-1991,Yuille-Kosowski-1994}.
Annealing methods are similar to (Markov chain) Monte Carlo methods,
which aim at sampling many points 
from a specific distribution
(i.e., for example at fixed temperature).
For Monte Carlo methods it is important to have (nearly) independent samples
and the correct limiting distribution for the Markov chain.
For annealing methods the aim is to find the correct minimum 
by smoothly changing the temperature from a finite value to zero.
For the latter it is thus less important to model the distribution
for nonzero temperatures exactly, but
it is important to use an adequate
cooling scheme for lowering the temperature.

\subsection{Hyperfields}
\label{hyperfields}

The hyperparameters $\theta$ 
considered so far have been real or integer {\it numbers}, 
or {\it vectors} with real or integer components $\theta_i$.
In this section we will discuss 
priors parameterized by functions,
called {\it hyperfields} \cite{Lemm-BFT-1999},
resulting in a still larger flexibility of the formalism.
In numerical calculations where functions have to be discretized
hyperfields stand for high dimensional hyperparameter vectors.

Using hyperfields 
one has to keep in mind
that a gain in flexibility at the same time
tends to lower the influence of the prior.
For example,
consider as hyperfield a completely adaptive reference potential 
$\theta(x)$ = $v_0(x)$
within a Gaussian prior (\ref{gaussprior}).
Then, for any $v(x)$
the prior energy vanishes 
for $v_0(x)$ = $v(x)$.
In the absence of additional hyperpriors $p(\theta)$
the corresponding MAP solution for the hyperfield 
$\theta(x)$ = $v_0(x)$
is thus
$\theta^*(x)$ = $v(x)$
for which the Gaussian prior (\ref{gaussprior})
becomes uniform in $v(x)$.
Hence the price to be paid for the additional flexibility
introduced by hyperfields
are weaker priors
and a large number of additional degrees of freedom.
This can considerably complicate calculations and
requires sufficiently restrictive hyperpriors for the hyperfields.

Let us define {\it local hyperfields} $\theta(x)$ 
to be  hyperfields depending on the position variable $x$. 
(In general hyperfields can be introduced 
which depend on other real variables or 
on several position variables.) 
Local hyperfields can be used, for example,
to adapt templates or inverse covariances locally.
To this end, 
we express real symmetric, positive (semi)\-definite inverse covariances
by square roots or (real) {\it filter operators} ${\bf W}$,
so that
\be
{\bf K}_0 = {\bf W}^T{\bf W}
.
\ee
In components
\be
{\bf K}_0(x,x^\prime) 
= \int \!dx^{\prime\prime}\; 
   {\bf W}^T(x,x^{\prime\prime}){\bf W}(x^{\prime\prime},x^{\prime})
,
\ee
and therefore
\bea
\mel{{v}-v_0}{{\bf K}_0}{{v}-v_0}
&=&
\int\! dx\,dx^\prime\, dx^{\prime\prime}\,
[{v}(x)-v_0(x)]
\nonumber\\
&&\times\;
{\bf W}^T(x,x^{\prime}){\bf W}(x^{\prime},x^{\prime\prime})
\nonumber\\
&&\times\;
[{v}(x^{\prime\prime})-v_0(x^{\prime\prime})]
\nonumber\\
&=&
\int \! dx\, |\omega (x)|^2
,
\eea
where we define the  {\it filtered difference}
\be
\omega (x)
=
\int \!dx^\prime \, {\bf W}(x,x^\prime) 
[{v}(x^\prime)-v_0(x^\prime )]
.
\label{filtered-diff}
\ee
For instance,
a square root (\ref{discrete2})
of the discrete negative Laplacian (\ref{discrete1})
corresponds for $v_0\equiv 0$ to a filtered difference
$\omega(x)$ = $v(x+1)-v(x)$.

The exponent of a Gaussian prior for a local potential ${v}$
can thus be written as an integral over $x$,
\be
p({v}) \propto e^{-E(v)}
;\quad
E(v) = \frac{1}{2}\int \!dx \, |\omega(x)|^2
.
\label{Gauss-omega}
\ee
In contrast to 
Eqs.~(\ref{integer-hyper-t}) and (\ref{integer-hyper-K})
the representation (\ref{Gauss-omega}) 
is well suited for introducing local hyperfields.
For instance,  
an adaptive prior 
\be
p({v}|\theta) = e^{-E(v|\theta)}
,
\label{hyper-prior}
\ee
with a real local hyperfield
$\theta(x)\in [0,1]$
can be obtained by
mixing locally two alternative filtered differences
\be
\omega (x;\theta) 
= [1-\theta(x)] \, \omega_1(x) + \theta(x) \,\omega_2(x)
\label{hyper-function-omega}
,
\ee
where the two $\omega_i$
may differ in their filters and/or reference potentials.
In that case 
the hyperfield $\theta(x)$ 
can locally select
the best mixture of the filtered differences
$\omega_i$, i.e., 
that one which yields in (\ref{hyper-prior})
the largest probability
or smallest prior energy 
\bea
E(v|\theta) 
&=& \frac{1}{2}
\int \!dx |\omega (x;\theta)|^2
+\ln Z_{\cal V}(\theta)
\label{local-hyper-p-r}
\\
&=&\frac{1}{2} \! \int \!\!dx 
 \Big| [1-\theta(x) ] \omega_1(x)
 +\theta(x) \omega_2(x)
 \Big|^2
\!+\ln Z_{\cal V}(\theta)
.
\nonumber
\eea
Here the normalization factor 
\be
Z_{\cal V}(\theta)
=
\int_{v\in {\cal V}} 
d \!v\, e^{-\frac{1}{2} \int \!dx |\omega (x;\theta)|^2}
,
\ee
depends in general on $\theta$
if the filters of the $\omega_i$ differ.
Clearly, 
allowing an unbounded $-\infty\le \theta(x)\le \infty$
any function $\omega (x;\theta)$
can be written in the form of Eq.~(\ref{hyper-function-omega}),
provided $\omega_1(x)\ne \omega_2(x)$ for all $x$.

In contrast to soft mixing with real functions $\theta(x)$
a binary local hyperfield $\theta(x)\in \{0,1\}$
implements hard switching 
between alternative filtered differences.
Since in the binary case
$\theta^2$ = $\theta$,
$(1-\theta)^2$ = $(1-\theta)$,
and 
$\theta(1-\theta)$ = $0$,
Eq.~(\ref{local-hyper-p-r})  
becomes [compare Eq.~(\ref{mix-prior-bin})]
\bea
E({v}|\theta) 
&=&
\frac{1}{2} \int \!dx \, 
 \Big( [1-\theta(x)]|\omega_1(x)|^2
\nonumber\\
&&
\quad +\theta(x) |\omega_2(x)|^2
 \Big)
+\ln Z_{\cal V}(\theta)
\label{local-hyper-p}
,
\eea
while for real $\theta(x)$
Eq.~(\ref{local-hyper-p-r})  
includes a mixed term in $\omega_1\omega_2$.
It is sometimes helpful to transform
an unrestricted real hyperfield $-\infty\le g(x)\le\infty$
into a bounded real hyperfield 
$\theta(x)\in [0,1]$ by
\be
\theta(x) = \sigma(g(x)-\vartheta)
,
\label{def-B}
\ee
with threshold $\vartheta$
and sigmoidal transformation
\be
\sigma(x) = \frac{1}{1+e^{-2\nu x}}
= \frac{1}{2} (\tanh(\nu x) + 1)
.
\label{sigmoid-bsp}
\ee
In the limit $\nu\rightarrow\infty$
the transformation $\sigma(x)$ of (\ref{sigmoid-bsp}) 
approaches the step function $\Theta(x)$ 
and (\ref{def-B}) results in a binary 
$\theta(x)$ = $\Theta(g(x)-\vartheta)\in \{0,1\}$.

Analogous to the global mixing or global switching
in Eq.~(\ref{integer-hyper-t})
and Eq.~(\ref{integer-hyper-K}),
the alternative filtered differences $\omega_i (x)$
at position $x$
in Eq.~(\ref{hyper-function-omega})
can be constructed by local mixing or switching 
between 
template functions 
$v_1(x^\prime)$, $v_2(x^\prime)$
or filters 
${\bf W}_1(x,x^\prime)$, ${\bf W}_2(x,x^\prime)$
using a local hyperfield $\theta(x)$,
\bea
v_x(x^\prime;\theta) 
&=& 
[1-\theta(x)] \, v_1(x^\prime) + \theta(x)\, v_2(x^\prime),
\label{hyper-function-t}
\\
{\bf W}(x,x^\prime; \theta) &=& 
[1\! -\theta(x)] {\bf W}_{1}(x,x^\prime)
\! + \theta(x) {\bf W}_{2}(x,x^\prime)
\label{hyper-function-W}
.\;\;
\eea
It is important to note that the local templates or 
reference potentials
$v_x(x^\prime; \theta)$ 
are functions 
of $x^\prime$ and $x$.
Indeed, to obtain a filtered difference $\omega(x;\theta)$ at position $x$,
a reference function $v_x$ is needed for all $x^\prime$ for which 
the corresponding ${\bf W}(x,x^\prime)$
is nonzero, since
\be
\omega(x;\theta )
=
\int\!dx^\prime\, 
{\bf W}(x,x^\prime) 
[{v}(x^\prime)-v_x(x^\prime;\theta)]
.
\ee
In this way the whole template function $v_x(x^\prime;\theta)$,
rather than individual function values $v_0(x,\theta)$, 
is adapted individually for every local filtered difference.
In particular, the local reference potentials of Eq.~(\ref{hyper-function-t})
have to be distinguished from one global,
locally adapted reference potential
\be
v_0(x^\prime;\theta) 
=
[1-\theta(x^\prime )] \, v_{1}(x^\prime)
 + \theta(x^\prime )\, v_{2}(x^\prime)
\label{mixing-t}
,
\ee
which at first glance seems to be the natural generalization of
Eq.~(\ref{integer-hyper-t}) to local hyperfields.
Only in Gaussian prior terms 
with the identity ${\bf I}$ as covariance,
local template functions $v_x(x^\prime, \theta)$
are not required.
In that case $v_{x}(x^\prime;\theta)$ 
is only needed for $x$ = $x^\prime$
and we may directly write
$v_{x}(x^\prime;\theta)$ 
=
$\tilde v_{0}(x^\prime;\theta)$,
skipping the variable $x$, and obtain the prior energy
\be
\frac{1}{2}\int\! dx\; |\omega(x;\theta)|^2
=
\frac{1}{2}\scp{v-\tilde v_0(\theta)}{v-\tilde v_0(\theta)}
.
\label{identity-cov}
\ee
We remark that one can also generalize
Eq.~(\ref{hyper-function-t}), 
which uses the same  
$v_1(x^\prime)$, $v_2(x^\prime)$ for all $x$,
by working with reference potentials 
$v_{1,x}(x^\prime)$, $v_{2,x}(x^\prime)$
which vary with the position $x$ 
at which the filtered difference $\omega(x)$
is required. This yields
\be
v_x(x^\prime;\theta) 
=
[1-\theta(x)] \, v_{1,x}(x^\prime)
+ \theta(x)\,  v_{2,x}(x^\prime)
.
\label{hyper-function-t-nonlocal}
\ee

For binary $\theta(x)$ 
Eq.~(\ref{hyper-function-W}) 
corresponds
to an inverse covariance
\bea
{\bf K}_0(\theta) 
&=& \int \!dx\; {\bf K}_x(\theta) 
=
\int \!dx \, {W}_{x}(\theta){W}^T_{x}(\theta)
\nonumber\\
&=& \int \!\! dx
\left(
[1-\theta(x)] {W}_{1,x}{W}^T_{1,x}
+  \theta(x)  {W}_{2,x}{W}^T_{2,x}
\right)
\qquad
\label{invcov}
\eea
with
\be
{\bf K}_x(\theta) = {W}_{x}(\theta){W}^T_{x}(\theta)
\ee
written as dyadic product of the vector
$W_{x}(\theta )$ = ${\bf W}(x,\cdot\,;\theta)$
and with analogously defined $W_{i,x}$ = ${\bf W}_i(x,\cdot)$.
For $\theta$--dependent inverse covariances
the normalization factors $Z_{\cal V}(\theta)$ become 
$\theta$--dependent. They have to be included
when integrating over $\theta$ or 
solving for the optimal $\theta$ in MAP.

In Eqs.~(\ref{hyper-function-t}) and (\ref{hyper-function-W})
it is straightforward to introduce
two binary hyperfields $\theta$, $\theta^\prime$,
one for the reference potential $v_x$ and one for 
the filter ${\bf W}$.
This results in a conditional prior
\bea
p({v}|\theta,\theta^\prime)
&\propto&
e^{-\frac{1}{2}
\int\!dx\, 
\mel{{v} - v_x(\theta)}{{\bf K}_x(\theta^\prime)}{{v}-v_x(\theta)}
}
\nonumber\\
&=&
e^{-\frac{1}{2} \int \!dx\, |\omega(x;\theta,\theta^\prime)|^2}
.
\eea
Here we can write
\bea
\int\! dx \, |\omega(x;\theta,\theta^\prime)|^2 
&=&
\mel{{v}-v_0(\theta,\theta^\prime)}
{{\bf K}_0(\theta^\prime)}{{v}-v_0(\theta,\theta^\prime)}
\nonumber\\
&&+
\int \!dx \,
\mel{v_x(\theta)}{{\bf K}_x(\theta^\prime)}{v_x(\theta)}
\nonumber\\
&&-
\mel{v_0(\theta,\theta^\prime)}
  {{\bf K}_0(\theta^\prime)}{v_0(\theta,\theta^\prime)}
,
\label{eff-E}
\eea
with an effective template $v_0(\theta,\theta^\prime)$
given by
\be
v_0(\theta ,\theta^\prime)
=
{\bf K}_0(\theta^\prime)^{-1}
\int\!dx\, {\bf K}_x(\theta^\prime ) \,  v_x(\theta)
,
\ee
and effective inverse covariance ${\bf K}_0(\theta^\prime)$
=
$\int \! dx\, {\bf K}_x(\theta^\prime)$
as in Eq. (\ref{invcov}).
Since 
the last two terms in Eq.~(\ref{eff-E}) are ${v}$--independent constants 
(only depending on $\theta$, $\theta^\prime$)
we see that for fixed hyperfields
this prior is minimized by
$v$ = $v_0(\theta,\theta^\prime)$.
For given hyperparameters $\theta$, $\theta^\prime$
we can write
$p({v}|\theta,\theta^\prime)\propto e^{-E({v}|\theta,\theta^\prime)}$
with a prior energy of the form
$E({v}|\theta,\theta^\prime)$
=
$\frac{1}{2}
\mel{{v}-v_0(\theta,\theta^\prime)}
    {{\bf K}_0(\theta^\prime )}{{v}-v_0(\theta,\theta^\prime)}$.

As the product of Gaussians is again a Gaussian
several Gaussian prior factors can easily be combined.
In this way one can implement a nonlocal property like smoothness 
and still avoid local template functions $v_x(x^\prime, \theta)$
by combining a Gaussian prior with ${\bf K}_0$ = ${\bf I}$
as in (\ref{identity-cov})
with a Gaussian prior with nondiagonal covariance and 
zero (or fixed) template,
\be
E({v}|\theta) =
\frac{1}{2}
\scp{{v}-\tilde v_0(\theta)}{{v}-\tilde v_0(\theta)}
+\frac{1}{2}
\mel{{v}}{{\bf K}}{{v}}
.
\label{local+laplace}
\ee
Combining both terms yields
\bea
E({v}|\theta) 
&=&
\frac{1}{2} 
\bigg( 
\mel{{v}-v_0(\theta)}{{\bf K}_0}{{v}-v_0(\theta)}
\nonumber\\
&&\quad  + \;\;
\mel{\tilde v_0(\theta)}
    {{\bf I}-{\bf K}_0^{-1}}{\tilde v_0(\theta)}
\bigg)
,
\label{local+laplace2}
\eea
with the second term 
being independent of $v$
and 
with effective template and effective inverse covariance
\be
v_0(\theta) = {\bf K}_0^{-1} \tilde v_0(\theta)
,\quad 
{\bf K}_0 = {\bf I}+ {\bf K}
.
\ee
For differential operators ${\bf K}_0$
the effective $v_0(\theta)$ 
is thus a smoothed version of $\tilde v_0(\theta)$.

The extreme case would be to treat
$v_0$ and ${\bf W}$ itself as unrestricted hyperfields.
As already discussed,
this just eliminates the corresponding prior term.
Hence, to restrict the flexibility,
typically a smoothness hyperprior may be imposed
to prevent highly oscillating functions $\theta (x)$.
For real $\theta(x)$, for example, a smoothness prior
like a Laplacian prior $\mel{\theta}{\! -\!\Delta}{\theta}/2$ can be used 
in regions where it is defined. 
(The space of functions
for which a smoothness prior 
with discontinuous templates is defined
depends on the locations of the discontinuities.)
An example of a non--Gaussian hyperprior is
\be
p(\theta) \propto 
e^{-\frac{\tau}{2} \int\!dx \, C_\theta(x)}
,
\label{hyperprior-C}
\ee
where $\tau$ is a constant 
and 
\be
C_\theta(x) = 
\sigma \left( \left(\frac{\partial \theta}{\partial x}\right)^2 
  - \vartheta_\theta\right)
,
\label{jumps}
\ee
with a sigmoid $\sigma(x)$ as in  (\ref{sigmoid-bsp}).
For $\nu\rightarrow\infty$ the sigmoid approaches a step function
and $C_\theta(x)$ 
becomes zero at locations where the square of the first derivative
is smaller than a certain 
threshold $0\le \vartheta_\theta < \infty$,
and one otherwise.
For discrete $x$ one can analogously 
count the number of jumps 
larger than a given threshold.
One can then penalize the number $N_d(\theta)$ 
of discontinuities
where 
$\left(\partial \theta/\partial x\right)^2$ = $\infty$
and use 
\be
p(\theta) \propto e^{-\frac{\tau}{2} N_d(\theta)}
.
\label{hyperprior-Nd}
\ee
In the case of a binary field
this corresponds 
to counting the number of times the field changes its value.
The expression $C_\theta$
of Eq. (\ref{jumps})
can be generalized to
\be
C_\theta(x)
= \sigma\left( |\omega_\theta(x)|^2-\vartheta_\theta\right)
,
\label{Cdef}
\ee
where,
analogous to Eq.~(\ref{filtered-diff}), 
\be
\omega_\theta(x)
=
\int \!dx^\prime \, 
{\bf W}_\theta(x,x^\prime) 
[\theta(x^\prime)-t_\theta(x^\prime)]
,
\ee
with template
$t_\theta(x^\prime)$
representing the expected form for the hyperfield,
and a filter operator 
${\bf W}_\theta$
defining a distance measure for hyperfields.
Parameters of the hyperprior like $\tau$ 
in Eq. (\ref{hyperprior-C}) or Eq.~(\ref{hyperprior-Nd})
can be treated as higher level hyperparameters.

\subsection{Non--Gaussian priors and auxiliary fields}
\label{Non--Gaussian-priors}

As an alternative to introducing hyperfields $\theta(x)$
one can work with priors which are
explicitly non--Gaussian with respect to $v$.
This can be done by introducing auxiliary fields
$B(x;v)$ 
whose function values are not considered 
as independent variables
but are directly defined as functionals of $v$.
(For the sake of simplicity
we will for $B(x;v)$ also write
$B(x)$ or $B(v)$, depending on the context.)
Like hyperfields,
auxiliary fields 
can select locally the best adapted filtered difference
from a set of alternative $\omega_i$.

For instance, consider the auxiliary field
[compare with Eqs.~(\ref{def-B}) or (\ref{Cdef})]
\be
B(x) = 
\sigma \left(u(x) - \vartheta \right)
,
\label{jumps2}
\ee
where
\be
u(x) = |\omega_1(x)|^2-|\omega_2(x)|^2
,
\ee
$\vartheta$ represents a threshold,
$\sigma (x)$ a sigmoidal function as in (\ref{sigmoid-bsp}),
and the $\omega_i$ are filtered differences 
defined in terms of $v$
according to Eq.~(\ref{filtered-diff}).
Again a binary field $B(x)$ is obtained 
by letting the sigmoid approach the step function.
Because the $\omega_i$ depend on $v$,
it is clear from the definition (\ref{jumps2})
that the auxiliary field $B(x)$ is no independent hyperfield
but has values being functionals of ${v}$.
Notice that $B(x)$ 
is nonlocal with respect to ${v}(x)$
if $\omega_i(x)$ is nonlocal;
a value $B(x)$ then depends 
on more than one ${v}(x)$--value.
For a negative Laplacian prior in one--dimension
Eq.~(\ref{jumps2}) reads,
\be
B(x) = 
\sigma \left( 
\left|\frac{\partial ({v}-v_1)}{\partial x}\right|^2
-\left|\frac{\partial ({v}-v_2)}{\partial x}\right|^2
            - \vartheta
\right)
.
\label{jumps2b}
\ee
While auxiliary fields $B(x)$ are directly determined by ${v}$,
hyperfields are indirectly coupled to $v$
through the MAP stationarity equations.
Conversely,
an auxiliary field $B(x)$ can be treated formally 
as independent hyperfield
if a Lagrange multiplier term 
\mbox{$\lambda
\left[
B(x)-\sigma \left(u(x) - \vartheta 
\right)
\right]$}
is added to the prior energy
in the limit $\lambda\rightarrow\infty$.

Like hyperfields $\theta(x)$
auxiliary fields $B(x)$
can be used 
to adapt reference potentials $v_0$ or filters ${\bf W}$.
However,
a prior as in  Eq.~(\ref{gaussprior})
is non--Gaussian with respect to $v$
if $v_0(B)$ and ${\bf K}_0(B)$ 
depend on $B$ and thus also on $v$.
Furthermore, analogous to hyperpriors $p(\theta)$,
additional prior terms $p(B(v))\propto \exp(-E_B(v))$ for $v$ 
can be included,
formulated in terms of an auxiliary field $B(x)$.
As in Eq.~(\ref{local-hyper-p})
a binary $B(x)$ can switch between two filtered differences
\be
|\omega(x;B)|^2
=
[1-B(x)] |\omega_1(x)|^2 
+
B(x) |\omega_2(x)|^2 
,
\label{binary-B}
\ee
within a (non--Gaussian) prior for ${v}$
\be
p({v}) \propto
e^{-E(v)-E_B(v)}
,
\label{b-prior}
\ee
where the normalization factor 
$Z$ = $\int \!dv\,e^{-E(v)-E_B(v)}$ 
of (\ref{b-prior})
is by definition independent of $v$. 
Hence it can be skipped for MAP calculations
also for non--Gaussian $p(v)$.
In Eq.~(\ref{b-prior})
\be
E(v) = \frac{1}{2} \int\!dx\, 
\left(
[1-B(x)] |\omega_1(x)|^2 
+
B(x) |\omega_2(x)|^2 
\right)
,
\label{omega-B-energy}
\ee
according to Eq.~(\ref{binary-B}),
while $E_B(v)$ depends on $v$
only through $B(v)$.
For example, the number of switchings
can be restricted by taking
\be
E_B(v) = \frac{\tau}{2}N_d(B)
,
\label{additional-b-prior}
\ee
where
$N_d(B)$ counts the number of discontinuities of $B(x)$.
Other choices, for real $B(x)$, 
are quadratic energies
\be
E_B(v) = \frac{\tau}{2}\int \!dx |\omega_B(x)|^2 
\label{quad-err}
\ee
or non--quadratic energies of the form
\be
E_B(v) = \frac{\tau}{2}\int \!dx \,C_B (x)
\label{non-quad-err}
\ee
where, similar to (\ref{Cdef}),
\be
C_B(x)
= \sigma\left( |\omega_B(x)|^2-\vartheta_B\right)
.
\label{cforb}
\ee
and
\be
\omega_B(x)
=
\int \!dx^\prime \, 
{\bf W}_B(x,x^\prime) 
[B(x^\prime)-t_B(x^\prime)]
,
\label{filtered-diff-ng}
\ee
is a filtered difference of $B$ 
with filter operator
${\bf W}_B$
and template
$t_B$.

Let us compare a non--Gaussian prior 
built of prior energies \ (\ref{omega-B-energy})
and (\ref{additional-b-prior}) 
for a binary auxiliary field (\ref{jumps2})
\be
p(v) \propto
e^{-\frac{1}{2} \int\!dx\, 
\left(
[1-B(x)] |\omega_1(x)|^2 
+
B(x) |\omega_2(x)|^2 
\right)
-\frac{\tau}{2} N_d(B)
}
,
\label{cmpB}
\ee
with the similar--looking
combination of Gaussian prior (\ref{local-hyper-p})
with hyperprior (\ref{hyperprior-Nd})
for a binary hyperfield,
\be
p({v},\theta) 
=p(v|\theta) p(\theta) 
\propto
\label{omega-theta-energy}
\label{cmpT}
\ee
\[
e^{-\frac{1}{2} \int\!dx\, 
\left[
(1-\theta(x)) |\omega_1(x)|^2 
+
\theta(x) |\omega_2(x)|^2 
\right]
-\frac{\tau}{2} N_d(\theta)
-\ln Z_{\cal V}(\theta)
}
.
\]
Eq.~(\ref{cmpT}) works with conditional probabilities $p(v|\theta)$, 
hence the corresponding 
normalization factors are in general $\theta$--dependent 
and have to be included
for MAP calculations.
Typically, MAP solutions 
for $B$, $N_d(B)$ and $C_B$
being directly defined in terms of the corresponding MAP solution for $v$
are different from the MAP solutions for $\theta$, 
$N_d(\theta)$ and $C_\theta$,
respectively.
However, if the filtered differences $\omega_i$ 
in Eq.~(\ref{omega-theta-energy})
differ only in their templates,
the normalization term can be skipped.
Then
assuming 
$\vartheta$ = $0$,
$p(\theta) \propto  1$,
$p(B) \propto 1$
the two equations are equivalent
for 
$\theta(x)$ = $\Theta\left(|\omega_1(x)|^2-|\omega_2(x)|^2\right)$.
In the absence of hyperpriors,
it is indeed easily seen 
that this is a selfconsistent solution for $\theta$
for every given ${v}$.
In general, however, when 
hyperpriors are included,
another solution for $\theta$ 
may have a larger posterior.

Hyperpriors $p(\theta)$ 
or additional auxiliary prior terms $p(B)$
can be useful to enforce specific
global constraints for $\theta(x)$ or $B(x)$.
In natural images, for example, discontinuities 
are expected to form closed curves.
Priors or hyperpriors, 
organizing discontinuities along lines or closed curves, 
are thus important for image segmentation or image restoration
\cite{Geman-Geman-1984,Poggio-Torre-Koch-1985,Marroquin-Mitter-Poggio-1987,Geiger-Girosi-1991,Zhu-Yuille-1996}.
A similar method has been used
in the determination of piecewise smooth relaxation time spectra
from rheological data
\cite{Roths-Maier-Friedrich-Marth-Honerkamp-2000}.

\begin{figure}
\vspace{-0.5cm}
\begin{center}
\epsfig{file=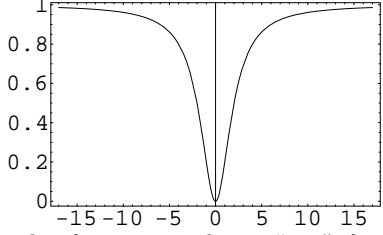, width=50mm}\\
\end{center}
\vspace{-0.5cm}
\caption{Example of a non--quadratic
``cup''--function
$\psi(x)$ = $a( 1.0 - 1/(1+(|x-x_0|/b)^\gamma))$, 
with
$a$= $5$,
$b$ = $10$,
$\gamma$ = $0.7$,
$x_0$ = $0$.
}
\label{Zhu-Mum-pic}
\end{figure}

Another  useful class of
non--Gaussian priors
generalizing (\ref{Gauss-omega})
has the form \cite{Winkler-1995,Zhu-Mumford-1997,Zhu-Wu-Mumford-1997}
\be
p(v)
\propto
e^{-\frac{1}{2} \int\! dx\, \psi[\omega(x)]}
,
\ee
where $\psi$ is a non--quadratic function.
This function $\psi$
can be fixed in advance for a given problem
or adapted using hyperparameters.
Typical choices to allow discontinuities
are symmetric ``cup'' functions 
with minimum at zero and flat tails 
for which one large step is cheaper than many small ones
(see Fig.~\ref{Zhu-Mum-pic}).

Table \ref{collection}
summarizes the basic variants of prior energies 
discussed in the paper.

\begin{table}[ht]
\begin{center}
\begin{tabular}{|c|c|}
\hline
\multicolumn{2}{|c|}{Gaussian prior\rule[-2mm]{0mm}{6mm}}\\
\hline\rule[-2mm]{0mm}{6mm}
$E(v)$ = 
$\frac{1}{2} \mel{v-v_0}{{\bf K}_0}{v-v_0}$ 
& (\ref{gaussprior}) \\
\hline
\multicolumn{2}{|c|}{with hyperparameter $\theta$\rule[-2mm]{0mm}{6mm}}\\
\hline\rule[-2mm]{0mm}{6mm}
$E(v|\theta)$ = 
$\frac{1-\theta}{2}\mel{v-v_1}{{\bf K}_1}{v-v_1}$ &\\
$\quad\qquad +\frac{\theta}{2}\mel{v-v_2}{{\bf K}_2}{v-v_2}$
\rule[-2mm]{0mm}{6mm}
&(\ref{mix-prior-bin})\\
\hline
\multicolumn{2}{|c|}{with local hyperfield $\theta(x)$\rule[-2mm]{0mm}{6mm}}\\
\hline\rule[-2mm]{0mm}{6mm}
$E(v|\theta)$ = 
$\frac{1}{2} \int \!dx \, 
 \Big( [1-\theta(x)]|\omega_1(x)|^2$ &\\
$\qquad\qquad+\theta(x) |\omega_2(x)|^2
 \Big)
+\ln Z_{\cal V}(\theta)$
& (\ref{local-hyper-p})\\
$E(v|\theta)$ = 
$\frac{1}{2}
\scp{{v}-\tilde v_0(\theta)}{{v}-\tilde v_0(\theta)}
+\frac{1}{2}
\mel{{v}}{{\bf K}}{{v}}$
\rule[-2mm]{0mm}{6mm}
&(\ref{local+laplace})\\
\hline
\multicolumn{2}{|c|}{
Non--Gaussian prior with auxiliary field $B(x;v)$\rule[-2mm]{0mm}{6mm}}\\
\hline\rule[-2mm]{0mm}{6mm}
$E(v)$ = 
$\frac{1}{2} \int\!dx\, 
\left(
[1-B(x)]|\omega_1(x)|^2 
+
B(x) |\omega_2(x)|^2 
\right)$ 
&(\ref{omega-B-energy})\\
\hline
\end{tabular}
\end{center}
\caption{Summary of basic prior energy variants discussed in this paper.}
\label{collection}
\end{table}

\section{Stationarity equations}
\label{stationarity-equations}

To reconstruct a local potential $v$
in MAP we have to maximize the posterior $p(v|D)$
with respect to $v$.
If the functional derivative of the posterior with respect to $v$ exists,
the reconstructed potential can be found by solving the 
stationarity equation
\be
\delta_{v} \ln p(v|D) = 0
,
\label{stat-eq}
\ee
where we have chosen the logarithm for technical convenience, and 
$\delta_{v}$ denotes the functional derivative with respect to $v$.

For observational data consisting of 
$n$ independent position measurements
the posterior (\ref{bayestheorem}) reads
\be
p(v|D) 
\propto 
p(v) \prod_{i=1}^n p(x_i|\hat x,v)
.
\label{posterior2}
\ee
To formulate the stationarity equation (\ref{stat-eq})
we have to calculate the functional derivatives 
of likelihood and prior.
For inverse quantum statistics \cite{Lemm-IQS-2000} 
the likelihood for position measurements (\ref{pos-likelihood})
on a canonical ensemble (\ref{canonical})
depends on the eigenfunctions and eigenvalues 
of the $v$--dependent Hamiltonian $H(v)$.
We thus have to find the functional derivatives
of the eigenfunctions $\phi_\alpha$ and eigenvalues $E_\alpha$.
Those can be obtained by taking the functional derivative
of the eigenvalue equation 
$H\ket{\phi_\alpha}$ = $E_\alpha \ket{\phi_\alpha}$,
where we will assume the eigenfunctions 
to be orthonormalized.
Choosing
$\scp{\phi_\alpha}{\delta_{v(x)}\phi_\alpha}$ = 0
and
utilizing  
\be
\delta_{v(x)} H (x^\prime,x^{\prime\prime})
= 
\delta_{v(x)} V (x^\prime,x^{\prime\prime})
= 
\delta(x-x^\prime) \delta (x^\prime-x^{\prime\prime})
,
\ee
we find for nondegenerate eigenfunctions
\bea
\delta_{v(x)} E_\alpha 
&=& \mel{\phi_\alpha}{\delta_{v(x)} H}{\phi_\alpha}
=|\phi_\alpha(x)|^2
,
\label{deltaE-nonp}
\\
\delta_{v(x)} \phi_\alpha(x^{\prime})
&=& \sum_{\gamma\ne \alpha} \frac{1}{E_\alpha-E_\gamma} 
\phi_\gamma(x^{\prime})\phi^*_\gamma(x) \phi_\alpha (x)
.
\eea
It follows for the functional derivative of the likelihood
\bea
\delta_{v(x)}p(x_i|\hat x,v)
&=&
\av{\left(\delta_{v(x)}\phi^*(x_i)\right) \phi (x_i)} 
\nonumber\\&&
+\av{\phi^*(x_i)\delta_{v(x)}\phi (x_i)}
\nonumber\\&&
-
\beta \Big(
\av{|\phi (x_i)|^2 |\phi (x)|^2}
\nonumber\\&&
-\av{|\phi (x_i)|^2}\av{|\phi (x)|^2}
\Big)
.
\label{der-like}
\eea

Having obtained Eq.~(\ref{der-like})
for the likelihood 
we now have to find the functional derivative of the prior.
For the Gaussian prior (\ref{gaussprior})
one gets directly
\be
\delta_{v} \ln p(v) 
= -{\bf K}_0(v-v_0)
.
\label{prior-dev}
\ee

If hyperparameters $\theta$ are included
and treated in MAP
(i.e., not integrated out by Monte Carlo techniques),
the posterior has to be maximized simultaneously
with respect to $v$ and $\theta$.
We have already mentioned that $\theta$--dependent inverse covariances
lead to normalization factors which are independent of $v$
but depend on $\theta$.
Such factors have to be included 
when maximizing with respect to $\theta$.

As a non--Gaussian example
consider a prior
where two filtered differences
are mixed by an auxiliary field $B(x)$
and an additional prior factor $p(B)$ is included,
for example to prevent fast oscillations of $B(x)$.
With $B(x)$ = $\sigma(u(x)-\vartheta)$,
threshold $\vartheta$,
sigmoidal function  $\sigma(x)$
as in Eq.~(\ref{sigmoid-bsp}),
and 
$u(x)$ = $|\omega_1(x)|^2-|\omega_2(x)|^2$
this gives
\be
p(v) \propto
e^{-\frac{1}{2} \, \int\! dx \, 
\big| [1-B(x)] \omega_1(x) + B(x) \omega_2(x) 
\big|^2 
-E_B
}
.
\label{non-gauss-prior}
\ee
Analogous to Eq.~(\ref{b-prior}),
the term 
\be
E_B = \int\!dx\, E_B(x)
,
\ee
represents an auxiliary prior energy
formulated in terms of the mixing function $B(x)$.
Like $\omega(x)$ the function value $E_B(x)$
may depend on the whole function $B$
and not necessarily only on the function value $B(x)$.
Using $\omega_i(x)$ = $\scp{x}{{\bf W}_i (v-v_i)}$
we find
\be
\delta_{v(x)} \omega_i(x^\prime) 
= {\bf W}_i(x^\prime,x)
,
\ee
and thus
\be
\delta_{v(x)} u(x^\prime)
=
2\left({\bf W}^T_1(x,x^\prime) \, \omega_1(x^\prime)
-{\bf W}^T_2(x,x^\prime) \, \omega_2(x^\prime)
\right).
\ee
Furthermore, we obtain for the functional derivative of $E_B$
\be
\delta_{v(x)} E_B(x^\prime)
=
\int\!dx^{\prime\prime}\,
\left[
\delta_{v(x)} B(x^{\prime\prime})
\right]
\, 
\left[
\delta_{B(x^{\prime\prime})} E_B(x^\prime)
\right]
,
\ee
where with Eq.~(\ref{jumps2})
\be
\delta_{v(x)} B(x^{\prime\prime})
=
\sigma^\prime(u(x^{\prime\prime})-\vartheta)
\delta_{v(x)} u(x^{\prime\prime})
,
\ee
and
$\sigma^\prime(u)$ = $d\sigma(u)/du$.
For a prior energy as in (\ref{quad-err}) which is quadratic in  $B(x)$
\be
E_B(x) = |\omega_B(x)|^2
,
\ee
$\omega_B(x)$ defined in Eq.~(\ref{filtered-diff-ng}),
the functional derivative with respect to $B(x)$ becomes
\be
\delta_{B(x)}E_B(x^\prime)
 = 
2 {\bf W}_B^T(x,x^\prime)\, \omega_B(x^\prime)
.
\ee
For a  non--Gaussian prior with energy (\ref{non-quad-err})
an additional derivative of the sigmoid appears.
Now all terms can be collected and inserted into the 
functional derivative 
of the prior (\ref{non-gauss-prior})
\bea
\delta_{v} \ln p(v)
&=&
-\int\! dx \, 
\Big(
\left[[1-B(x)] \omega_1(x) + B(x) \omega_2(x) 
\right] 
\nonumber\\&&
\qquad\times
\big(
[1-B(x)] \delta_v\omega_1(x) 
+ B(x) \delta_v\omega_2(x) 
\nonumber\\&&
\qquad\qquad
+\;\delta_v B(x) [\omega_2(x)-\omega_1(x)] 
\big)
\nonumber\\&&
\qquad+\;\delta_v E_B(x) \Big)
.
\eea

The Bayesian approach to inverse quantum mechanics 
is not restricted to position measurements,
but allows to deal with all kinds of observations
for which the likelihood can be calculated.
To have better information about the depth of a potential
it is useful to include information on the
ground state energy of a system.
For instance,
including a noisy  measurement of the average energy
\be
U 
= \av{E}
=
\sum_\alpha p_\alpha E_\alpha
,
\ee
yields an additional factor in the posterior of the form
\be
p_U \propto e^{-E_U}
,\quad
E_U =
\frac{\mu}{2} (U - \kappa)^2
\label{averageE-penal}
.
\ee
In the noise free limit 
$\mu\rightarrow\infty$
this yields $U\rightarrow\kappa$.

Calculating 
the functional derivative of $U$ 
with respect to a local potential
\be
\delta_{v(x)} U =
\av{\delta_{v(x)} E}-\beta \av{E\; \delta_{v(x)} E}
+ \beta \av{E} \av{\delta_{v(x)} E}
,
\ee
it is straightforward to obtain
\be
\delta_{v(x)} E_U 
=
  \mu\left(U\!-\!\kappa\right)
\av{|\phi (x)|^2\left[1-\beta \left( E  -U \right) \right]}
.
\ee

Stationarity equations are typically nonlinear 
and have to be solved by iteration.
A possible iteration scheme is 
\bea
v^{(r+1)}
&=&
v^{(r)}\! +
\eta {\bf A}^{-1}
\Big[\delta_v \ln p(v^{(r)})
\label{iter1}
\nonumber\\&&
\quad +\sum_i \delta_{v}\ln p(x_i|\hat x,v^{(r)})
-\delta_{v} E_U^{(r)}
\Big]
.
\label{iteration}
\eea
Here $\eta$ is a step width which can be optimized
by a line search algorithm
and 
the positive definite operator ${\bf A}$ 
distinguishes different learning algorithms.

\section{Numerical examples}
\label{numerical}

As numerical application of BIQM
and to test several variants 
of implementing {\it a priori} information
we will study the reconstruction of an approximately
periodic, one--di\-mensional potential.
Such a potential may represent a one--dimensional surface
where a periodic structure, 
e.g. that of a regular crystal, 
is distorted by impurities,
located at unknown positions and of unknown form.

To test the quality of reconstruction algorithms,
artificial data will be sampled
from a model with known ``true'' potential $v_{\rm true}$.
Selecting a specific prior model
and applying the corresponding Bayesian reconstruction algorithm
to the sampled data, 
we will be able to compare the reconstructed
potential with the original one.
In particular, we will take as true potential
the following perturbed periodic potential
\be
v_{\rm true}(x) = 
\left\{
{
\sin \left( \frac{2\pi}{6}x\right); \quad 1\le x\le12,\,\;  25\le x\le 36,
\atop
\!\!\!\!\!\!
\!\!\!\!\!\!\!\!\!\!\!\!
\!\!\!\!\!\!\!\!\!\!\!\!
\!\!\!\!
\sin \left( \frac{2\pi}{12}x\right); \quad 13\le x\le 24,
}
\right.
\ee
using for the numerical calculations a mesh 
of size 36.
Considering a system prepared as canonical ensemble
the potential $v_{\rm true}$
defines a corresponding canonical density operator $\rho$ 
as given in Eq.~(\ref{canonical}).
Artificial data $D$ can then be sampled
according to the likelihood model 
of quantum mechanics (\ref{qm-likelihood}).
For the following examples, $n$ = 200 data points representing 
position measurements have been sampled
using the transformation method
\cite{Press-Teukolsky-Vetterling-Flannery-1992}.
In all calculations we used periodic boundary conditions 
for quantum mechanical wave functions
while the potential $v$ has been set to zero at the boundaries.

\begin{figure}
\begin{center}
\epsfig{file=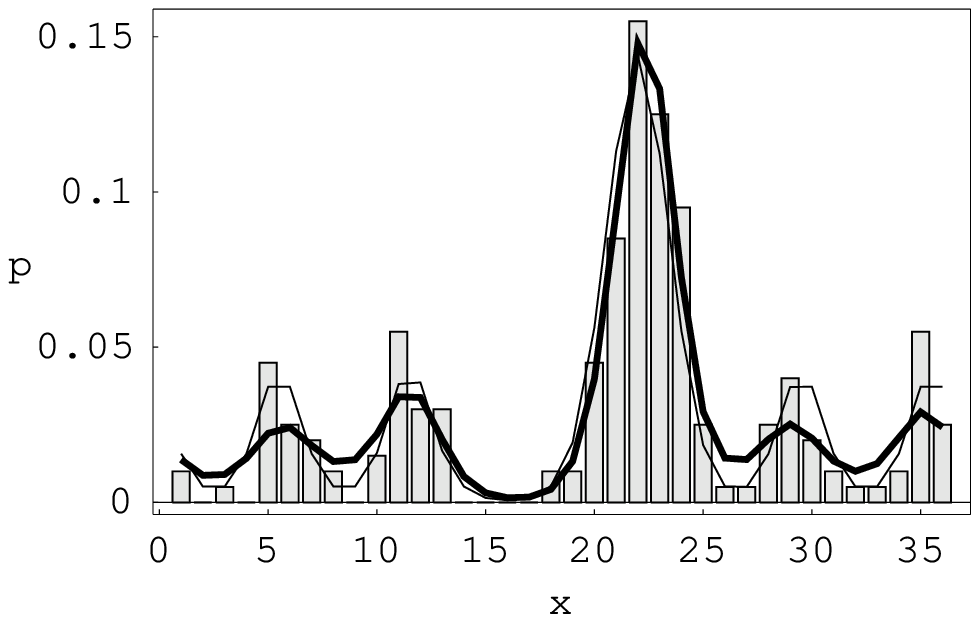, width= 67mm}
\epsfig{file=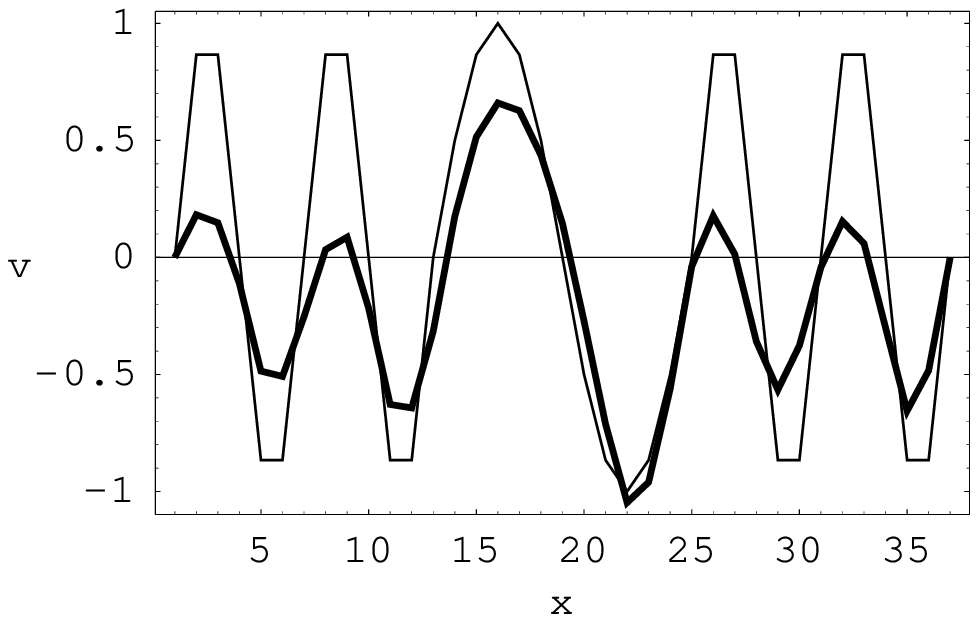, width= 67mm}
\end{center}
\caption{
Gaussian prior with Laplacian inverse covariance,
zero reference potential,
and additional noisy energy measurement.
Top: 
Empirical density $p_{\rm emp}$ (bars),
true likelihood $p_{\rm true}$ (thin), 
reconstructed likelihood $p_{\rm BIQM}$ (thick)
Bottom: 
Reconstructed potential $v_{\rm BIQM}$ (thick)
and true potential $v_{\rm true}$ (thin).
(With 200 data points,
$m$ = 0.25 for $\hbar$ = 1, $\beta$ = 4.
Gaussian prior (\ref{gaussprior})
with inverse covariance ${\bf K}_0$ = $-\lambda \Delta$,
$\lambda$ = 0.2, 
zero reference potential $v_0\equiv 0$,
and an additional energy penalty term of the form (\ref{averageE-penal})
with $\mu$ = 1000 
and $\kappa$ = $-0.330$,
equal to the true average energy $U(v_{\rm true})$.
The solution has been obtained by iterating according to 
Eq. (\ref{iteration}) with ${\bf A}$ = ${\bf K}_0$, starting 
with initial guess $v^{(0)} \equiv 0$.
The optimal step width $\eta$ has 
been determined for each iteration by a line search algorithm.)
}
\label{p162}
\end{figure}

We will now discuss the results of a Bayesian reconstruction
under varying prior models.
As first example, consider a simple Gaussian prior
(\ref{gaussprior}) 
with negative Laplacian inverse covariance
${\bf K}_0$ = $-\lambda \Delta$,
zero reference potential $v_0\equiv 0$,
and an additional prior factor (\ref{averageE-penal})
representing a noisy measurement of the average energy.
The reconstruction results
are shown in Fig.~ \ref{p162}.
In particular, the figure on top compares
the reconstructed likelihood 
$p_{\rm BIQM}(x|\hat x,v_{\rm BIQM})$
with the true likelihood
$p_{\rm true}(x|\hat x,v_{\rm true})$
and with the empirical density, i.e.,
the relative frequencies of the sampled data
\be
p_{\rm emp}(x) =
\frac{1}{n} \sum_{i=1}^n \delta(x-x_i)
.
\ee
Similarly, the lower figure compares
the reconstructed  potential $v_{\rm BIQM}$
with the true potential $v_{\rm true}$. 
Since information on the average energy
was available the depth of the potential
is well approximated at least at one of its minima.
This is sufficient to fulfill the noisy average energy condition.
However, because only smoothness and  no
periodicity information is implemented by the prior
the reconstructed potential is too flat.
The effect is stronger near the maxima 
than near the minima of the potential
because near the maxima only few data points are available
and hence the reconstructed potential is there dominated by the zero reference
potential in the smoothness prior.

\begin{figure}
\begin{center}
\epsfig{file=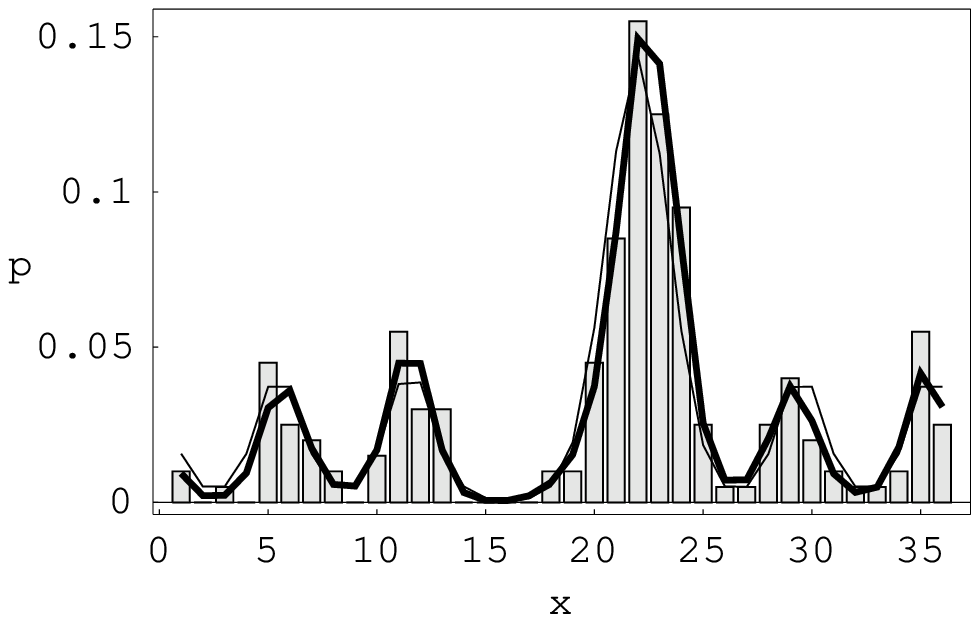, width= 67mm}
\epsfig{file=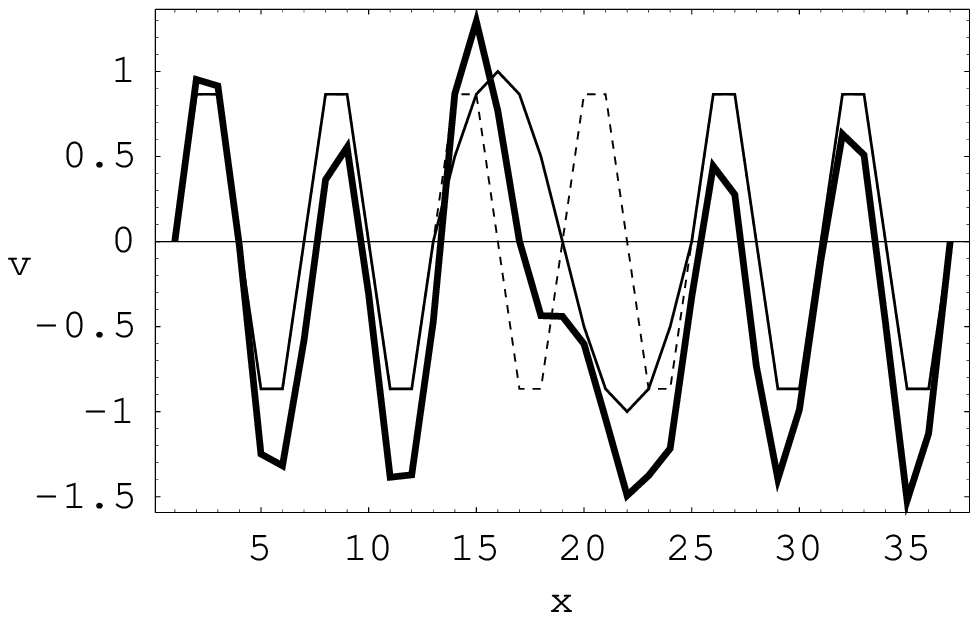, width= 67mm}
\end{center}
\caption{
Gaussian prior with periodic reference potential
without noisy energy measurement.
Top: 
Empirical density $p_{\rm emp}$ (bars),
true likelihood $p_{\rm true}$ (thin), 
reconstructed likelihood $p_{\rm BIQM}$ (thick)
Bottom: 
Reconstructed potential $v_{\rm BIQM}$ (thick).
true potential $v_{\rm true}$ (thin),  
and reference potential $v_0$ (dashed) 
of Eq.~(\ref{per-ref-prior}). 
(Number of data points
and parameters $m$, $\beta$,
${\bf K}_0$,
and $\lambda$
as for Fig.~\ref{p162} 
but with $\mu$ = 0.
The solution has been obtained by 
iterating according to (\ref{iteration})
as described for Fig.\ref{p162}
with initial guess
$v^{(0)} = v_0$.)
}
\label{p19}
\end{figure}

To include information on approximate periodicity
we have replaced in the next example
the zero reference potential $v_0\equiv 0$
by the strictly periodic reference potential
\be
v_0(x) = \sin \left( \frac{2\pi}{6}x\right)
,
\label{per-ref-prior}
\ee
shown as dashed line 
in the following figures of potentials.
A reconstruction 
with the periodic reference potential (\ref{per-ref-prior})
but without average energy information,
and starting the iteration with the reference potential as
initial guess $v^{(0)}$ = $v_0$
is shown in Fig.~\ref{p19}.
Due to missing average energy information
the depth of the potential is not well approximated.
It is also clearly visible in Fig.~\ref{p19}
that the smoothness prior
does not favor solutions which are 
similar to the reference $v_0$ itself 
but solutions
which have derivatives similar to that of $v_0$.
Fig.~\ref{p19} also displays
that the reconstruction of the potential does clearly identify
the impurity.
As the reference potential is not adapted
to the impurity region the reconstruction is there poorer 
than in the regular region.

Furthermore, it is worth emphasizing 
that the reconstructed likelihood 
fits the empirical density well,
even slightly better than the true likelihood does.
This is due to the flexibility of a nonparametric approach
which allows to fit the fluctuations of the empirical density 
caused by the finite sample size. 
The effect is well known in empirical learning and 
leads to so called ``overfitting''
if the influence of the prior becomes to small.
Since observational data 
influence the reconstruction only through the likelihood,
the reconstruction of potentials is in general 
a more difficult task than the reconstruction of likelihoods.
This indicates the special 
importance of {\it a priori} information 
when reconstructing potentials.
Indeed, even if the complete likelihood is given,
the problem of determining the potential
can still be ill--defined
in regions where the likelihood is small
\cite{Zhu-Rabitz-1999}.

A prior model with periodic reference potential 
can be made more flexible
by adapting 
amplitude, frequency, and phase
of the reference potential (\ref{per-ref-prior}).
For this purpose one can introduce a hyperparameter vector
$\theta$ = $(\theta_1,\theta_2,\theta_3)$
parameterizing amplitude, frequency, and phase
and take as reference potential
\be
v_0(x;\theta) 
= \theta_1 \sin \left( \frac{2\pi}{\theta_2}x+\theta_3\right)
.
\ee
The corresponding maximization of the posterior
with respect to $\theta$ is easy in that case
and does not change the results of Fig.~\ref{p19}
where the hyperparameters are already optimally adapted.

\begin{figure}
\begin{center}
\epsfig{file=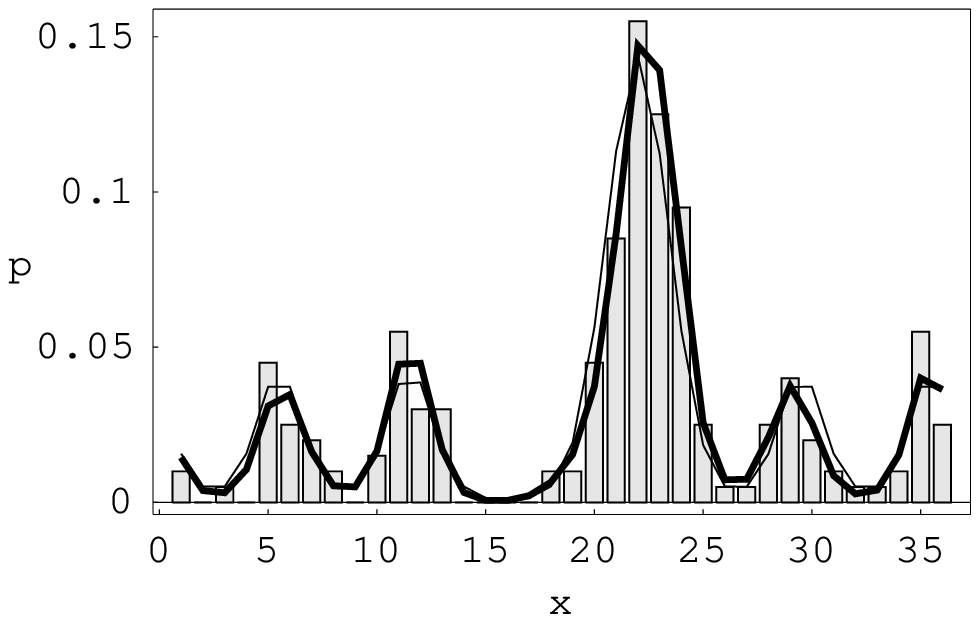, width= 67mm}
\epsfig{file=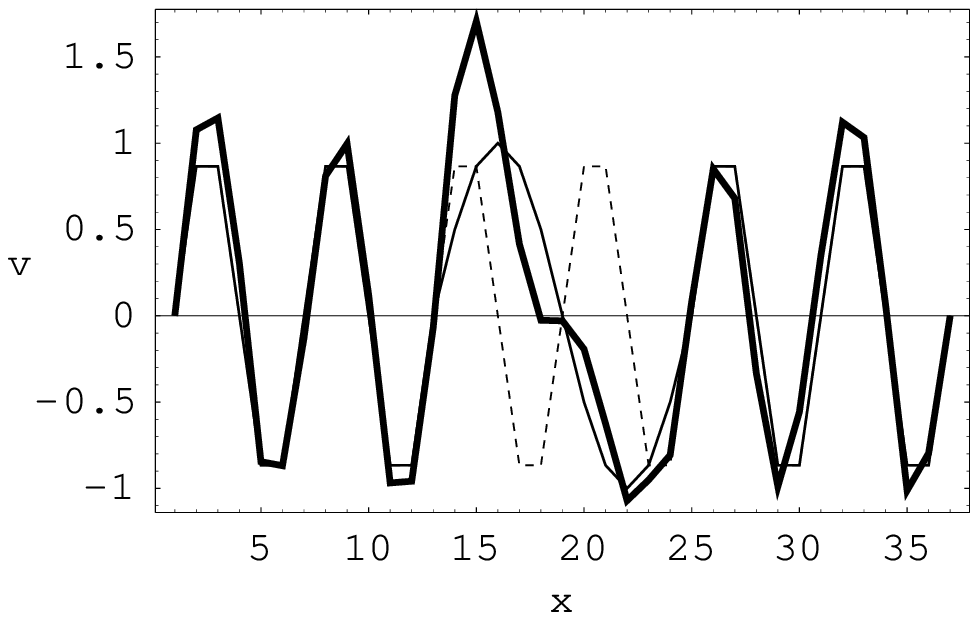, width= 67mm}
\end{center}
\caption{
Gaussian prior with periodic reference potential 
and additional energy measurement,
improving the approximation of the minima.
(Reference potential $v_0$ given in (\ref{per-ref-prior}),
energy penalty term as in (\ref{averageE-penal})
with $\mu$ = 1000
and 
$\kappa$ = $-0.330$.
All other parameters as for Fig.~\ref{p19}.
Iterated with
the solution shown in Fig.~\ref{p19} 
as initial guess $v^{(0)}$.)
}
\label{p22}
\end{figure}

\begin{figure}
\begin{center}
\epsfig{file=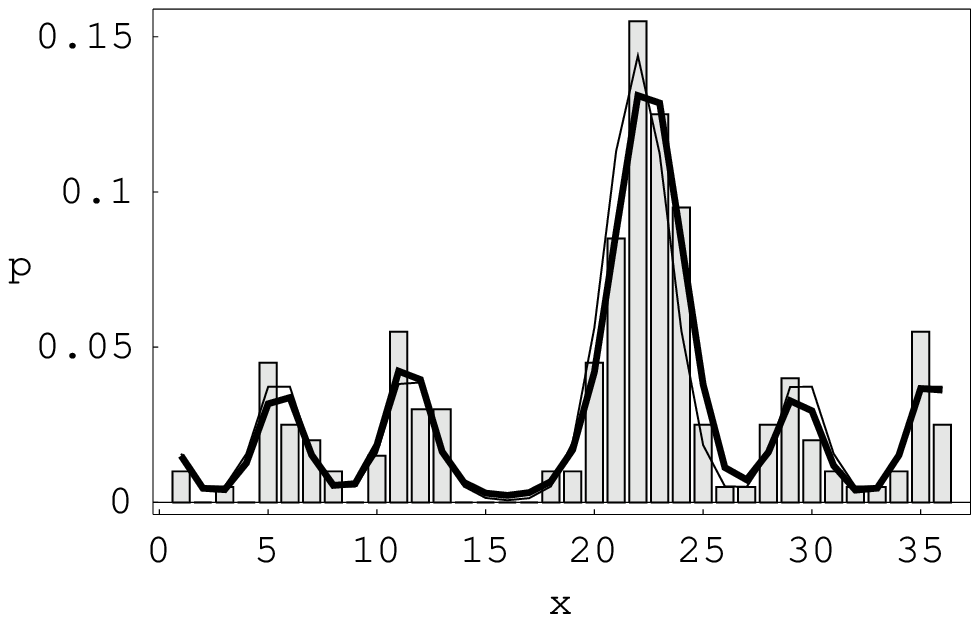, width= 67mm}
\epsfig{file=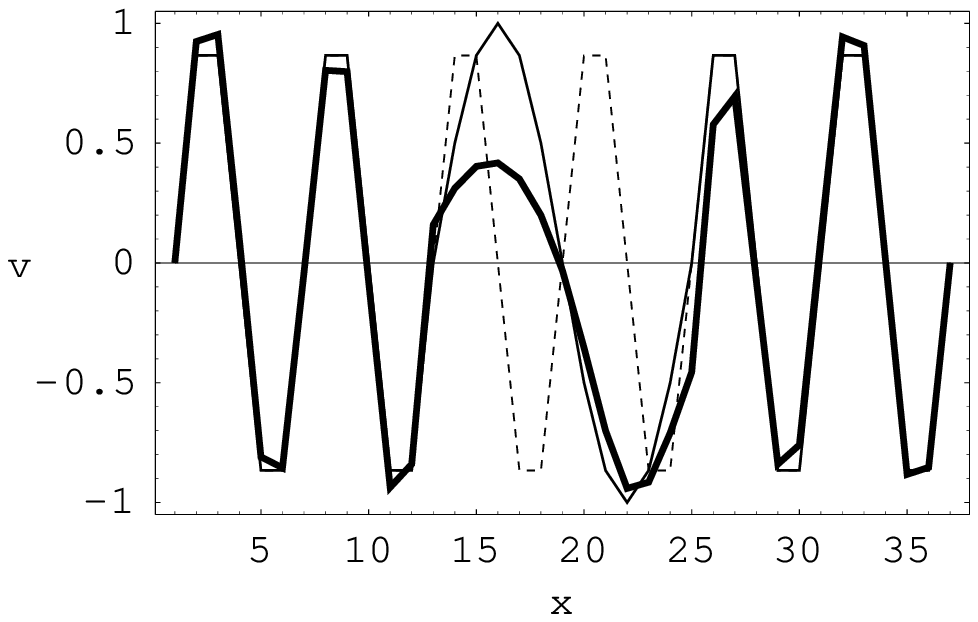, width= 67mm}
\end{center}
\caption{
Gaussian prior with periodic reference potential 
and additional energy measurement,
with initial guess $v^{(0)}$ different from that of Fig.~\ref{p22}.
(Reference potential $v_0$ given in (\ref{per-ref-prior}),
energy penalty term as in (\ref{averageE-penal})
All  parameters as for Fig.~\ref{p22}.
Iterated with
initial guess $v^{(0)}(x)$ = $v_0(x)$
for $0<x\le12,\,  25\le x$ and 
$v^{(0)}(x)$ = $0$ for $13\le x\le 24$.)
}
\label{p155}
\end{figure}

Including an additional noisy energy measurement
(\ref{averageE-penal})
Fig.~\ref{p22} shows that 
the depth of the 
potential is indeed better approximated
than in Fig.~\ref{p19}.
To avoid local maxima of the posterior
the solution of Fig.~\ref{p19} 
has been used as initial guess
and the factor $\mu$ multiplying the average energy term
has been slowly increased to its final value.
Fig.~\ref{p22} still only represents a
local and no global maximum of the posterior, 
as can be by seen by starting 
with a different initial guess $v^{(0)}$.
In Fig.~\ref{p155}
a better solution for the same parameters
is presented 
where the initial guess has been selected
using {\it a priori} information
about the location of the impurity region.

\begin{figure}
\begin{center}
\epsfig{file=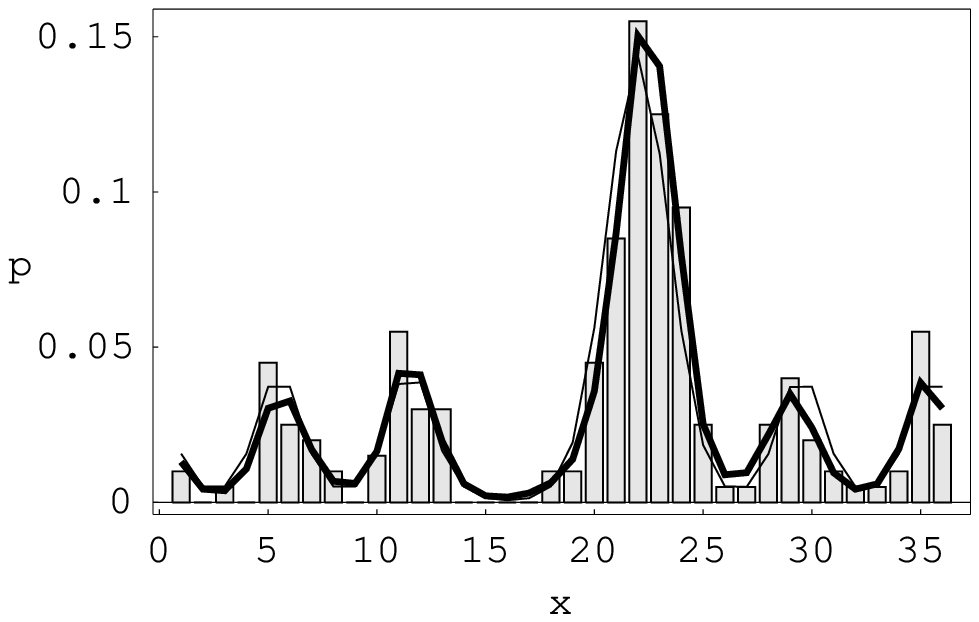, width= 67mm}
\epsfig{file=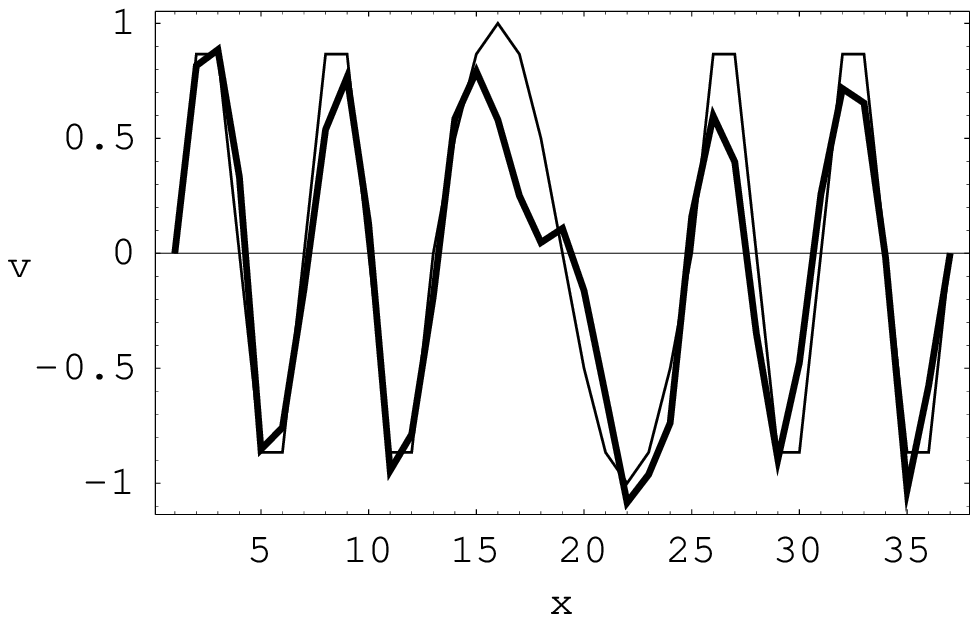, width= 67mm}
\end{center}
\caption{
Approximate periodicity implemented by an inverse covariance
${\bf K}_0$ 
=
$- \lambda (\Delta+\gamma \Delta_\theta)$
as in Eq.~(\ref{periodic-cov}).
(With $\gamma$ = 1.0, $\lambda$ = 0.2, 
a fixed $\theta$ = 6,
energy penalty term with $\mu$ = 1000,
and zero reference potential $v_0\equiv 0$.
Initial guess $v^{(0)}$ = $v_0\equiv 0$.
All other parameters as for Fig.~\ref{p19}.
%
)
}
\label{p31}
\end{figure}

Alternatively to a Gaussian prior with periodic reference,
approximate periodicity can be enforced by 
the inverse covariance of a Gaussian prior.
In this case the prior 
favors periodicity but no special form of the potential.
The prior is thus less specific
than a prior with explicit periodic reference function.
Corresponding BIQM results 
for the inverse covariance (\ref{periodic-cov})
are shown in 
Fig.~\ref{p31}. 
Indeed while the potential is well approximated 
in regions where many observations have been collected,
it is not as well approximated in regions where no or only few data 
are available.
These are the regions where the prior dominates the observational data.
In particular, in the case presented in Fig.~\ref{p31},
the zero reference function $v_0\equiv 0$
of an additional Laplacian smoothness prior
implements a tendency to flat potentials.

\begin{figure}
\begin{center}
\epsfig{file=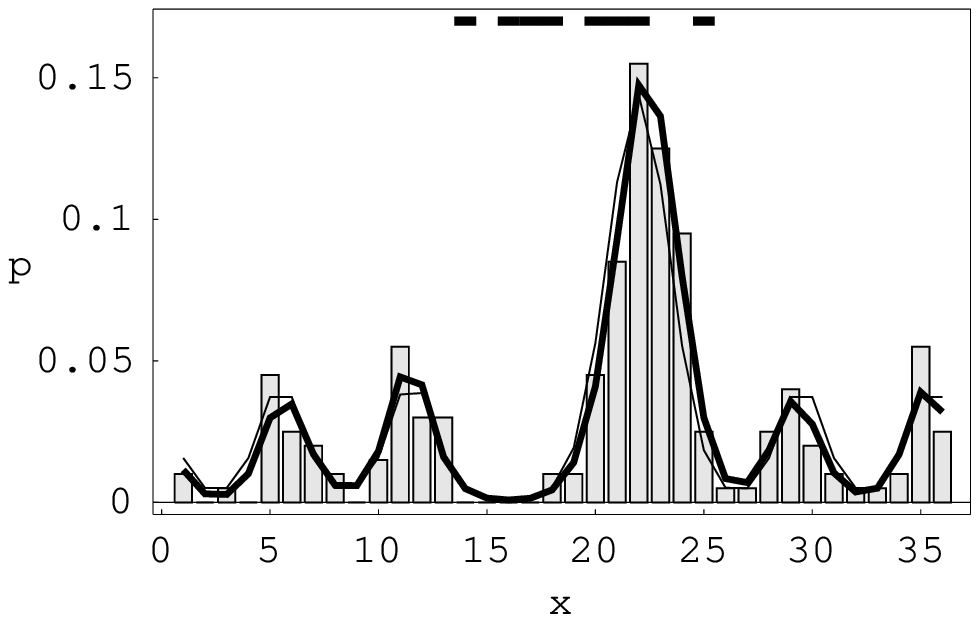, width= 67mm}
\epsfig{file=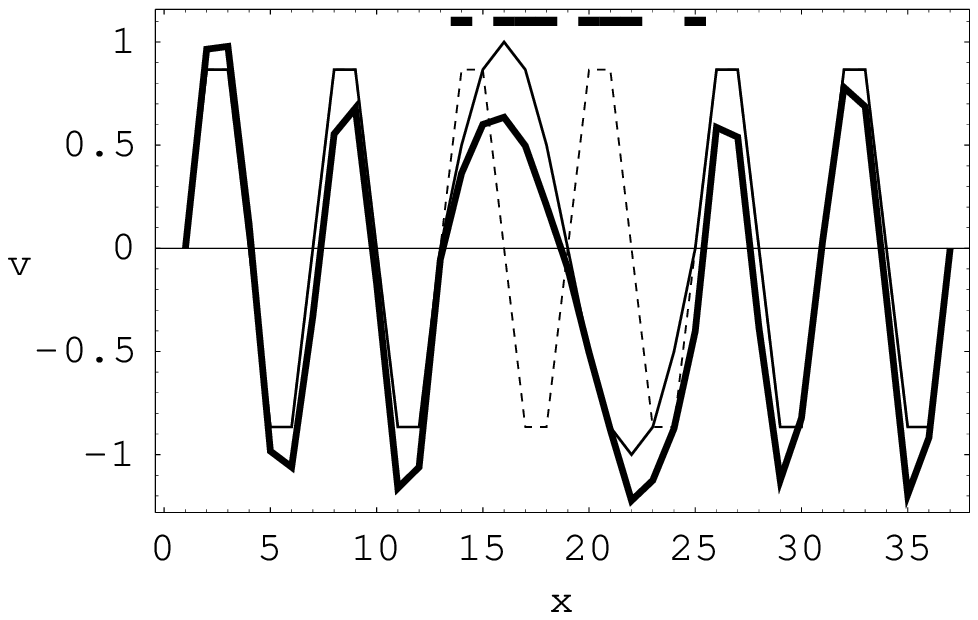, width= 67mm}
\end{center}
\caption{
Local switching between periodic 
and zero reference potential.
The black bars on top indicate regions where $B(x)$ = 1,
i.e., regions where impurities have been identified.
(Prior of Eq.\ (\ref{eq1}) with
$\lambda_1$ = 0.2, $\lambda_2$ = 0.2, 
$\mu$ = 0,
and reference potential as in (\ref{per-ref-prior}).
The $v$--dependent function $B(x)$ was slowly changed 
from a sigmoid to a step function
during iteration, keeping  the threshold $\vartheta$ = $0.15$ fixed.
All other parameters as
in Fig.~\ref{p19}.
Initial guess $v^{(0)}$ as for Fig.~\ref{p155})
}
\label{p102}
\end{figure}

If impurities are expected,
a prior with one fixed periodic reference potential
for the whole region is no adequate choice.
Near impurities one would like to 
switch off the standard periodic reference potential 
which in these regions will be misleading.
Because it is usually not known in advance 
where a given reference should be used and where not,
those regions  must be identified
during learning.
As first example we study a prior energy
similar to Eq.~(\ref{local+laplace}),
\bea
E(v) 
&=&
\frac{\lambda_1}{2}
\int\!dx\,
|v(x)-v_0(x)|^2 [1-B(x)]
 -\frac{\lambda_2}{2}\mel{v}{\Delta}{v},
\nonumber\\&&
\label{eq1}
\eea
which allows to switch off a given reference locally
by means of a binary switching function
defined as
$B(x)$ = $\Theta\left(|v(x)-v_0(x)|^2-\vartheta\right)$.
(An average energy term 
$E_U$ = $\frac{\mu}{2}(U-\kappa)^2$
could easily be included.)
In the prior energy (\ref{eq1}) the reference $v_0$ is only used
if $|v(x)-v_0(x)|^2$ is smaller than the given threshold $\vartheta$.
Starting with a smoothed version of Eq.~(\ref{eq1}) 
with a real mixing function 
$B(x)$ = $\sigma\left(|v(x)-v_0(x)|^2-\vartheta\right)$,
the results of Fig.~\ref{p102}
have been obtained by changing during iteration
$\sigma(x)$ slowly from a sigmoid to a step function.
Using a step function for $B$ directly from the beginning
leads to nearly indistinguishable results.
Compared to Fig.~\ref{p31}
the reconstruction in Fig.\ \ref{p102}
is improved mainly in the unperturbed region
where the algorithm can now use the correct reference potential.
An additional advantage is 
that the final auxiliary field $B(x)$
directly shows the identified impurity regions. 
One sees in  Fig.~\ref{p102}
that the auxiliary field $B(x)$ is always switched off
if the solution $v(x)$ is similar enough to the template $v_0(x)$.

\begin{figure}[ht]
\begin{center}
\setlength{\unitlength}{0.9mm}
\hspace{3cm}
\begin{picture}(90,20)
\thicklines
\put(0,0){\line(1,0){40}}  
\put(50,0){\line(1,0){10}}  
\put(80,0){\line(1,0){10}}  
\put(60,0){\line(0,1){10}} 
\put(80,0){\line(0,1){10}} 
\put(60,10){\line(1,0){20}}   
\put(10,0){\circle*{1.2}}  
\put(30,0){\circle*{1.2}}  
\put(60,0){\circle*{1.2}}  
\put(80,0){\circle*{1.2}}  
\put(41,5){\vector(1,0){8}} 
\put(49,5){\vector(-1,0){8}} 
\end{picture}
\end{center}

\begin{center}
\setlength{\unitlength}{0.9mm}
\begin{picture}(90,20)
\thicklines
\put(0,0){\line(1,0){10}}  
\put(10,0){\line(0,1){10}}  
\put(10,10){\line(1,0){30}}  
\put(50,0){\line(1,0){30}}  
\put(80,0){\line(0,1){10}}   
\put(80,10){\line(1,0){10}}   
\put(10,0){\circle*{1.2}}  
\put(30,0){\circle*{1.2}}  
\put(60,0){\circle*{1.2}}  
\put(80,0){\circle*{1.2}}  
\put(41,5){\vector(1,0){8}} 
\put(49,5){\vector(-1,0){8}} 
\end{picture}
\end{center}

\begin{center}
\setlength{\unitlength}{0.9mm}
\begin{picture}(90,20)
\thicklines
\put(0,0){\line(1,0){20}}  
\put(20,0){\line(0,1){10}}  
\put(20,10){\line(1,0){20}}  
\put(50,0){\line(1,0){10}}  
\put(60,0){\line(0,1){10}}  
\put(60,10){\line(1,0){10}} 
\put(70,0){\line(0,1){10}} 
\put(70,0){\line(1,0){10}}   
\put(80,0){\line(0,1){10}}   
\put(80,10){\line(1,0){10}}   
\put(10,0){\circle*{1.2}}  
\put(30,0){\circle*{1.2}}  
\put(60,0){\circle*{1.2}}  
\put(80,0){\circle*{1.2}}  
\put(41,5){\vector(1,0){8}} 
\put(49,5){\vector(-1,0){8}} 
\end{picture}
\end{center}
\caption{Generation of new trial configurations
for simulated annealing
by selecting two points randomly
and exchanging the values zero and one of the binary function in between.
This mechanism has been used to optimize the binary functions 
$c(x)$ and $\theta(x)$.}
\label{trial}
\end{figure}
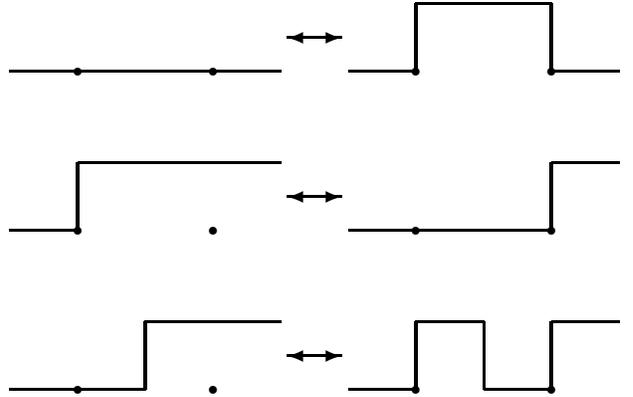

The two $v$--dependent terms in
Eq.~(\ref{eq1}) can be combined
[compare Eqs.~(\ref{local+laplace}) and (\ref{local+laplace2})].
Skipping a term which only depends
on $v$ through $B(x)$,
one arrives at another
prior which also implements local switching.
More general, choosing the prior energy (\ref{omega-B-energy}) 
for switching between
two filtered differences 
with two reference potentials  $v_1$ and $v_2$
leads to
\bea
E(v) 
&=&
 \frac{\lambda_1}{2}
\int\!dx\,
[1-B(x)] |\omega_1(x)|^2 
\nonumber\\&&
+
\frac{\lambda_2}{2}
\int\!dx\,
B(x) |\omega_2(x)|^2 
\label{eq2}
,
\eea
where the switching is controlled by
the binary function $B(x)$
=
$\Theta\left(|\omega_1(x)|^2-|\omega_2(x)|^2 -\vartheta\right)$
defined in terms of the filtered differences
$\omega_i(x)$ = $(\partial/\partial x) [v(x)-v_i(x)]$. 
A prior energy (\ref{eq2})
with two different nonzero reference potentials
$v_1$ and $v_2$ is obtained, for example,
when a different nonzero reference potential is given
for the unperturbed and the perturbed region.
The number of changes
in the switching function
$B(x)$ = $\Theta\left(|\omega_1(x)|^2-|\omega_2(x)|^2-\vartheta\right)$,
can be controlled
by adding a prior term $p(B)$
penalizing the number of times the function $B(x)$
changes its value.
To avoid local minima for binary $B(x)$,
simulated annealing techniques
are useful.
We have obtained an initial guess for $v$, 
and thus for $B(x)$,
by writing $v(x)$ = $[1-c(x)]v_1(x)+c(x)v_2(x)$
and optimizing the binary function $c(x)$
by simulated annealing
with respect to the likelihood and the additional prior $p(B)$.
In particular, starting from $c(x)$ = $0$,
new trial functions have been generated
by selecting two points $x_1$, $x_2$ randomly
and exchanging the function values zero and one
in between (see Fig.~\ref{trial}).
A new trial function has been accepted or rejected
using the Metropolis rule
$p($accept) = min$[1,\exp({-\beta_{\rm ann} \Delta E_{\rm ann}})]$
with $\Delta E_{\rm ann}$ denoting the difference in the error 
between actual function and new trial function.
In the present case we have
$E_{\rm ann}(v)$ = $\sum_{i} E(x_i|\hat x,v)$ + $E_B(v)$
where
$E(x_i|\hat x,v)$ = $- \ln p(x_i|\hat x,v)$ and
$p(B)\propto \exp(-E_B)$.
The annealing temperature $1/\beta_{\rm ann}$
decreases during optimization.

\begin{figure}
\begin{center}
\epsfig{file=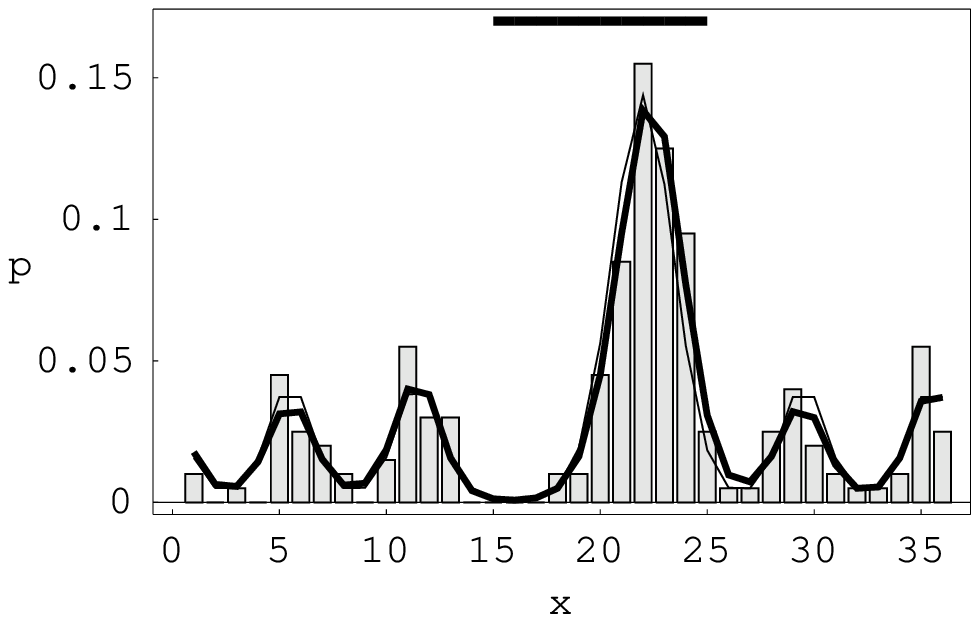, width= 67mm}
\epsfig{file=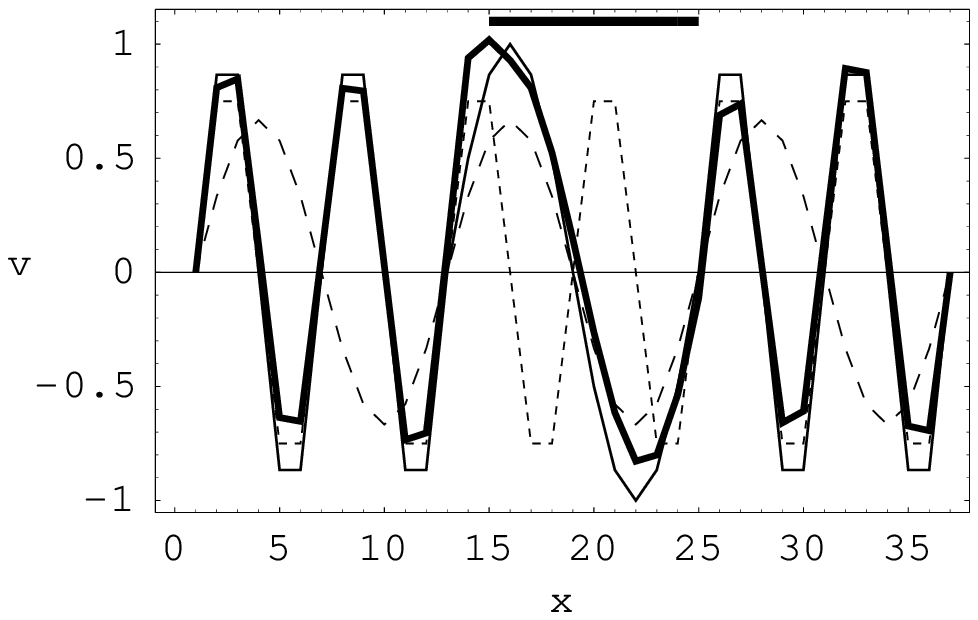, width= 67mm}
\end{center}
\caption{
Local switching between two nonzero reference potentials.
(Reference potentials $v_1$, $v_2$
given in 
(\ref{two-ref-potentialsA})
and
(\ref{two-ref-potentialsB}).
Prior of Eq.\ (\ref{eq2}),
with 
$\lambda_1$ = $\lambda_2$ = 10, $\mu$ = 0.
Step function for $B(x)$ with $\vartheta$ = 0. 
An additional prior $p(B)$ on $B$ has been included 
with $-\ln p(B)/10$
counting the number of discontinuities of the function $B(x)$.
Other parameters as
in Fig.~\ref{p19}.)
}
\label{p75}
\end{figure}

Fig.~\ref{p75}
shows the reconstruction results
using the following two reference potentials
\bea
v_1(x) &=& \frac{2}{3} \sin \left( \frac{2\pi}{6}x\right)
,
\label{two-ref-potentialsA}
\\
v_2(x) &=& \sin^2 \left( \frac{2\pi}{6}x\right)
{\rm sign}\left[\sin \left( \frac{2\pi}{6}x\right)\right]
.
\label{two-ref-potentialsB}
\eea
Compared to Fig.\ \ref{p102}
the reconstruction is improved
in the perturbed region,
where the algorithm can now rely on a useful reference potential.

\begin{figure}
\begin{center}
\epsfig{file=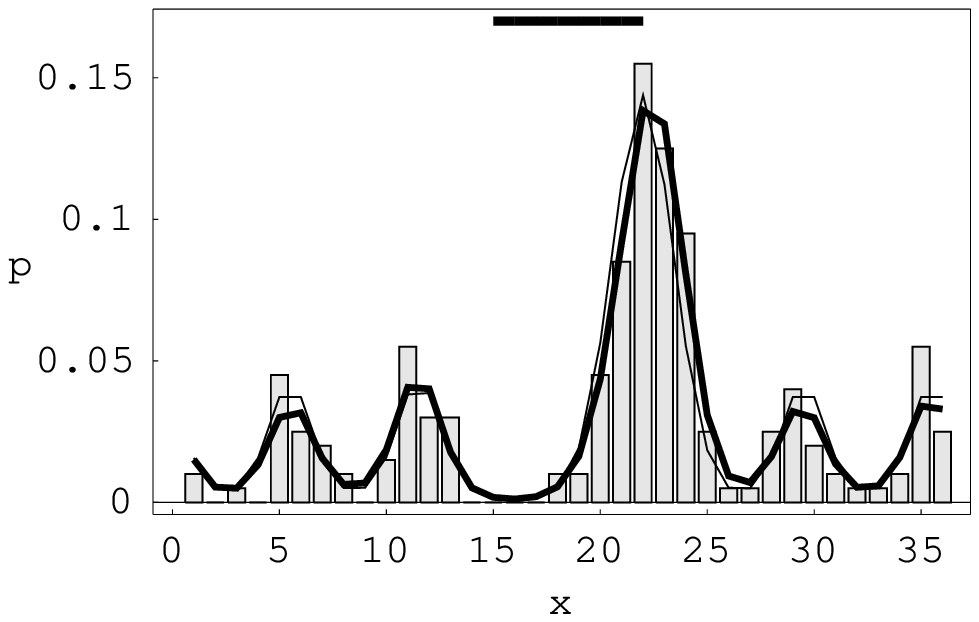, width= 67mm}
\epsfig{file=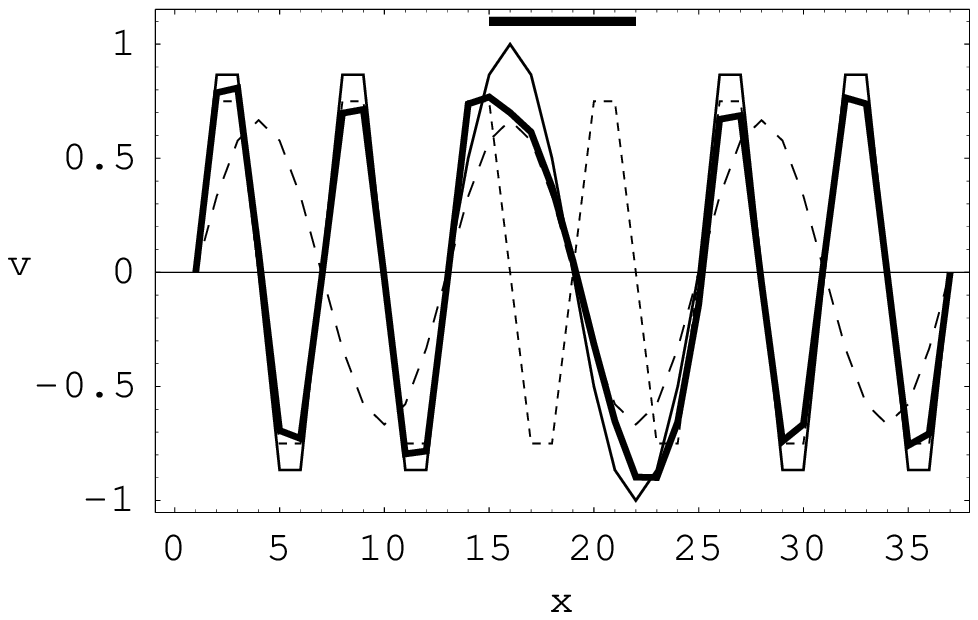, width= 67mm}
\end{center}
\caption{
Prior with local hyperfield.
(Prior of Eq.\ (\ref{eq3}),
with 
$\lambda_1$ = 10, $\lambda_2$ = 1,
$\vartheta$ = 0, $\mu$ = 0,
including a hyperprior $p(\theta)$ with 
$E_B/10$
counting the number of discontinuities of the hyperfield $\theta(x)$.
Other parameters as
in Fig.~\ref{p19}.)
}
\label{p120}
\end{figure}

Finally, the switching function can be introduced
as local hyperfield.
As an example for a prior with hyperfield,
Fig. \ref{p120} shows the reconstruction
with the prior energy
\be
E(v,\theta) 
=
\frac{\lambda_1}{2}\scp{v-v_0(\theta)}{v-v_0(\theta)}
-\frac{\lambda_2}{2}\mel{v}{\Delta}{v}
-\ln p(\theta)
\label{eq3}
,
\ee
where 
$v_0(x;\theta)$ = $v_1(x)[1-\theta(x)]+v_2(x)\theta(x)$
with the reference potentials of 
Eq.~(\ref{two-ref-potentialsA})
and Eq.~(\ref{two-ref-potentialsB}).
A hyperprior $p(\theta)$ has been used
penalizing the number of discontinuities of the hyperfield $\theta(x)$,
analogous to $p(B)$ for Fig.~\ref{p75}.
The $E(v|\theta)$ part of the prior energy (\ref{eq3}) 
is of the form (\ref{local-hyper-p}) with $\theta$--independent
covariances. Hence the $\theta$--independent normalization factor
can be skipped.
An initial guess 
for the  local hyperfield $\theta (x)$ has been obtained
by simulated annealing as described for Fig.~\ref{p75}.
As in this case optimization is required only
with respect to the $\theta$--dependent parts of the posterior,
optimizing $\theta(x)$ for given $v$ 
is faster than optimizing $v$ through $c(x)$
which requires diagonalization of the hamiltonian $H$ 
for every new trial function.
However, as $\theta(x)$ is independent of $v$, the hyperfield 
has to be updated during iteration which has also been done
by simulated annealing.
As expected, a reconstruction with the non--Gaussian prior
corresponding to the prior energy (\ref{eq2})
is very similar to a reconstruction 
using hyperfields as in  Eq.~(\ref{eq3}).

\section{Conclusion}

A nonparametric Bayesian approach
has been developed and applied 
to the inverse problem of reconstructing potentials
of quantum systems from observational data.
Relying on observational data only
the problem is typically ill--defined.
It is therefore essential
to include adequate {\it a priori} information.
Since reconstructed potentials
obtained by Bayesian Inverse Quantum Mechanics (BIQM)
depend sensitively on the implemented {\it a priori} information,
flexible prior models are required
which can be adapted to the specific situation under study.
In particular, the use of hyperparameters, hyperfields,
and non--Gaussian priors with auxiliary fields
has been discussed in detail.
In this paper we have focussed on
the implementation of approximate periodicity
for potentials in inverse problems of quantum statistics.
The presented prior models, however, can be useful 
for many empirical learning problems, 
including for example regression or general density estimation.
Several variants of implementing  {\it a priori} information
on approximate periodicity
have been tested and compared numerically.



\end{document}